\documentclass[twoside]{article}
\usepackage{a4}
\usepackage{tabularx}
\usepackage{epsf}
\usepackage{graphics}

\usepackage{amssymb}
\def\aj{AJ}%
\def\apj{ApJ}%
\def\apjs{ApJS}%
\def\aap{A\&A}%
\def\nat{Nature}%
\def\pasp{PASP}%

\def\bref{\vspace{4pt}\noindent\hangindent=10mm}

\begin{document}

\setcounter{figure}{0}
\setcounter{section}{0}
\setcounter{equation}{0}

\begin{center}
{\Large\bf
Stellar evolution of massive stars\\[0.2cm]
at very low metallicities}\\[0.7cm]

R. Hirschi, C. Fr\"ohlich, M. Liebend\"orfer and F.-K. Thielemann\\[0.17cm]
 Dept. of Physics and Astronomy, University of Basel\\
Klingelbergstr. 82, 4056 Basel, Switzerland\\
raphael.hirschi\@unibas.ch
\end{center}

\vspace{0.5cm}

\begin{abstract}
\noindent{\it
Recently, measurements of abundances in extremely metal poor (EMP) stars
have brought new constraints on stellar evolution models. Indeed, these
stars are believed to have been enriched only by one or a few stars.
The abundances observed in EMP stars can therefore be almost directly
compared with the yields of very metal poor or metal free stars. This
removes the complex filter of chemical evolution models. 
In an attempt
to explain the origin of the abundances observed, we computed
pre--supernova evolution models, explosion models and the
related nucleosynthesis.
In this paper, we start by presenting the 
pre-SN models of rotating single stars with metallicities ranging 
from solar metallicity down to almost
metal free. We then review key processes in core-collapse and bounce,
before we integrate them in a simplistic parameterization for 3D
MHD models,
which are well underway and allow one to follow the evolution of the
magnetic fields during collapse and bounce.
Finally, we present explosive nucleosynthesis results including neutrino
interactions with matter, which are calculated using the outputs of the
explosion models.

The main results of the pre-SN models are the following. First, 
primary nitrogen is produced in large amount in models with an initial
metallicity $Z=10^{-8}$. This large production is due to mixing of
carbon and oxygen in the H-burning shell and the main production occurs
between core helium and carbon burnings. 
Second, at the same metallicity of $Z=10^{-8}$
and for models with an initial mass larger than about 60 $M_\odot$,
rotating models may experience heavy mass loss (up to more than half of
the initial mass of the star). Several solar masses are lost during the
main sequence when the star reaches break-up velocities. The largest
amount of mass is lost during the red supergiant stage after rotational
mixing and convection have enriched the surface in CNO elements and
therefore increased its metal content. 
The chemical composition of these winds can qualitatively reproduce the
abundance patterns observed at the surface of carbon-rich EMP stars.
Third, our models predict an upturn of C/O at very low metallicities. 
Explosive nucleosynthesis including neutrino-matter interactions produce 
improved abundances for iron group elements, in particular for scandium 
and zinc. It also opens the way to a new
neutrino and proton rich process ($\nu$p-process) able to contribute to
the nucleosynthesis of elements with A$>$64.
}
\end{abstract}

\section{Introduction}
Massive stars ($M\gtrsim 10 M_\odot$) play an important role in
astrophysics. They are the progenitors of blue supergiants (BSG), red
supergiants (RSG), Wolf-Rayet (WR) and luminous blue variable (LBV)
stars.  At the end of their life, they explode as type II or Ib,c
supernovae (SN) and maybe also as long soft gamma-ray bursts (GRB). Their
cores after collapse become neutron stars (NS) or black holes (BH). They
are one of the main sites for nucleosynthesis, which takes place during
both pre-SN (hydrostatic) burnings as well as during explosive burnings.
A weak s-process occurs during H-burning and
r-process probably occurs during the explosion. 
Radioactive isotopes like $^{26}$Al and $^{60}$Fe detected by the INTEGRAL
satellite are produced by massive stars.
Massive stars even though they are much less numerous than low mass
stars contribute significantly to the integrated luminosity of galaxies.

At high redshifts ($z$, or low metallicities $Z$), 
their importance grows. The first
stars formed are thought to be all massive or even very massive (Bromm
2005) and
to be the cause of the re-ionisation of the universe. Finally, recent surveys of metal poor halo
stars provide many constraints for the early chemical evolution of our
Galaxy (Beers et al. 1992, Beers 1999, Israelian et al. 2004, 
Christlieb et al. 2004, Cayrel et al. 2004, Spite et al. 2005).

The measured abundances of many elements present a very small scatter
down to very low metallicities ([Fe/H]$\sim -4.2$, see Cayrel et al.
2004). This tends to prove that the interstellar medium
(ISM) was already well mixed at an early stage. However, r-process
elements show a scatter in their abundances contradicting this
trend (Ryan...).
There is no significant enrichment by pair-instability supernovae (PISN,
see Heger and Woosley 2002, Umeda and Nomoto 2002, 2003 for more 
details).
Production of primary nitrogen in massive stars (that died before the
formation of the halo metal poor star) is necessary in order to explain
the high abundance of nitrogen observed.

About a quarter of the extremely metal poor (EMP) stars show a strong 
enrichment of carbon (carbon-rich, CEMP), with [C/Fe]$\sim$2--4, where  
$[{\rm A/B}]= {\rm log_{10}}(X_{\rm A}/X_{\rm B})-
{\rm log_{10}}(X_{\rm A}/X_{\rm B})_\odot$. 
The CEMP stars can probably be divided into two subclasses, the Ba-rich
and the Ba-normal stars (see Ryan et al. 2005 and references therein).
According to Ryan et al. (2005), the Ba-rich CEMP stars have accreted
enriched matter from an AGB stars in a binary system in the course of their life.
On the other hand the Ba-normal CEMP stars would have formed from a cloud
already enriched in carbon and neutron capture elements. The cloud
itself would have been enriched by wind or supernova ejecta from massive
stars. 
The two most metal poor halo stars found up to now, HE0107-5240 
([Fe/H]=-5.3, see Christlieb et al. 2004)
and HE1327-2326 ([Fe/H]=-5.5, see Frebel et al. 2005 and Aoki et al
2005) belong to this last group, the Ba-normal CEMP stars. These two
stars both present strong but different enrichment in nitrogen and 
oxygen.

In the light of these recent observational results
and following the exploratory work of 
Meynet et al (2005), we explore the impact
that rotation can have in massive very low metallicity stars. In Sect.
2, we present the pre-SN models, their evolution and stellar yields for different
initial masses, rotation velocities and metallicities and compare them 
with EMP stars. In
Sect. 3, we describe the recent update in the explosion models and the
corresponding nucleosynthesis in 1D and the latest development of a new
3D MHD model. In Sect. 4 we give our conclusions.

\section{Pre-SN models}
\subsection{Description of the stellar models}
The computer model used to calculate the stellar models
 is described in detail in Hirschi et al
 (2004).
Convective stability is determined by the 
Schwarzschild criterion. 
Convection is treated as a diffusive process from oxygen burning
onwards.
The overshooting parameter is 0.1 H$_{\rm{P}}$ 
for H and He--burning cores 
and 0 otherwise. Instabilities induced by rotation
taken into account are meridional circulation 
and secular and dynamical shears. 
The reaction rates are taken from the
NACRE (Angulo et al 1999) compilation for the experimental rates
and from the NACRE website (http://pntpm.ulb.ac.be/nacre.htm) 
for the theoretical ones. 

At low metallicities the initial chemical composition
is calculated in the following way.
For a given metallicity $Z$ (in mass fraction), 
the initial helium mass fraction
$Y$ is given by the relation $Y= Y_p + \Delta Y/\Delta Z \cdot Z$, 
where $Y_p$ is the primordial
helium abundance and $\Delta Y/\Delta Z$ the slope of 
the helium--to--metal enrichment law. 
$Y_p$ = 0.24 and $\Delta Y/\Delta Z$ = 2.5 were used according to recent
determinations (see Izotov \& Thuan 2004 for
example).
For the mixture of the heavy elements, 
we adopted the same mixture as the one
used to compute the opacity tables for Weiss 95's alpha--enriched 
composition (Iglesias \& Rogers 1996). 

The mass loss rates are described and
discussed in Meynet, Ekstr\"om \& Maeder (2005). 
The mass loss rates (and opacities) are rather well 
determined for chemical compositions which are similar to solar (or
alpha--enriched mixing) composition or similar to a fraction of the
solar composition. However, very little was known about the mass
loss of very low metallicity stars with a strong enrichment in CNO
elements until recently. Vink \& de
Koter (2005) study the case of WR stars but a
crucial case, which has not been studied in detail yet, is the case of red
supergiant stars (RSG). As we shall see later, due to rotational and
convective mixing, the surface of the star is strongly enriched in CNO
elements during the RSG stage. Awaiting for future studies, 
it is implicitly assumed 
in this work (as in Meynet, Ekstr\"om \& Maeder 2005) 
that CNO elements have a significant contribution to
opacities and mass loss rates. 
Therefore the mass loss rates depend 
on metallicity as $\dot{M} \sim (Z/Z_{\odot})^{0.5}$, where
$Z$ is the mass fraction of heavy elements at the surface
of the star, also when the iron group elements content is much lower 
the CNO elements content. 

A specific treatment for mass loss was applied at break-up 
(see Meynet, Ekstr\"om \& Maeder 2005).
At break-up, the mass loss rate adjusts itself in such a way that an
equilibrium is reached between the envelope extension and the removal 
of gravitationally unbound mass.
In practice, however, since the critical limit contains mathematical
singularities, we considered that during the break-up phase, 
the mass loss rates should be such
that the model stays near a constant fraction (around 0.95) 
of the limit.

\subsection{Characteristics of the models}\label{vini}
The value of 300 km\,s$^{-1}$ used for the initial rotation velocity 
at solar metallicity
corresponds to an average velocity of about 220\,km\,s$^{-1}$ on the Main
Sequence (MS) which is
very close to the average observed value (see for instance Fukuda 1982). 
It is unfortunately not possible to measure the rotational velocity of very low
metallicity stars since they all died a long time ago.
Nevertheless, there is indirect evidence that stars with a lower metallicity have a
higher rotation velocity. Observational evidence is the higher ratio of Be to B stars in
the Magellanic clouds compared to our Galaxy (Maeder et al 1999). This can be due to the
difficulty of evacuating angular momentum during the star formation, which is even more
important at lower metallicities 
(see Abel et al 2002 for simulations of star formation in the early Universe).
Finally, a low metallicity star containing the same angular momentum
as a solar metallicity star has a higher surface rotation velocity due to
its smaller radius (one quarter of $Z_\odot$ radius for 20 $M_\odot$
stars). It is therefore worth studying models with an initial rotational 
velocities faster than 300 km\,s$^{-1}$. 

In order to compare the
models at different metallicities and with different initial masses
with another quantity than the surface velocity, 
the ratio $\upsilon_{\rm ini}/ \upsilon_{\rm crit}$ is used (see Table
\ref{table1}).
$\upsilon_{\rm crit}^2=GM/R_{\rm eb}(1- \Gamma)$, where $R_{\rm eb}$
is the equatorial radius at break--up and $\Gamma$ is the ratio of the
luminosity to the Eddington luminosity.
$\upsilon_{\rm ini}/ \upsilon_{\rm crit}$ 
 increases only as $r^{-1/2}$ for
models with the same angular momentum ($J$) but lower metallicity, 
whereas the surface rotational velocity increases as $r^{-1}$ 
($J\sim \upsilon r$).
The angular momentum could be used as well but it varies significantly
for models of different initial masses.
Finally,
$\upsilon_{\rm ini}/ \upsilon_{\rm crit}$ is a good indicator for the
impact of rotation on mass loss.

In the first series of models, the aim is to scan the parameter space of
rotation and metallicity with 20 $M_\odot$ models since a 20 $M_\odot$
star is not far from the average massive star concerning stellar yields. 
For this series,
two initial rotational velocities were 
used at very low metallicities. 
The first one is the same as at solar metallicity, 
300\,km\,s$^{-1}$.
The ratio $\upsilon_{\rm ini}/ \upsilon_{\rm crit}$ decreases with
metallicity (see Table \ref{table1} for the 20 $M_\odot$ models) 
for the initial velocity of 300\,km\,s$^{-1}$.
The second $\upsilon_{\rm ini}$ is 
500\,km\,s$^{-1}$ at Z=10$^{-5}$ ([Fe/H]$\sim $-3.6) and 600\,km\,s$^{-1}$
at Z=10$^{-8}$ ([Fe/H]$\sim$-6.6). These values have
ratios of the initial velocity to the break--up velocity, 
$\upsilon_{\rm ini}/\upsilon_c$ around 0.55, which is only slightly
larger than the solar metallicity value (0.44).

The 20 $M_\odot$ model at Z=10$^{-8}$
 and with 600\,km\,s$^{-1}$ has a total initial angular momentum $J_{\rm
tot}=3.3\,10^{52}$\,erg\,s which is the same as for 
of the solar metallicity 20 $M_\odot$ model with 300\,km\,s$^{-1}$ ($J_{\rm
tot}=3.6\,10^{52}$\,erg\,s).
So a velocity of 
600\,km\,s$^{-1}$, which at first sight seems extremely fast, is probably
the average velocity at Z=10$^{-8}$.

In the second series of models, we follow the exploratory work of
Meynet, Ekstr\"om \& Maeder (2005) and compute models at 
Z=10$^{-8}$ with initial masses of
40, 60 and 85 $M_\odot$ and initial rotational velocities of 700, 800 and
800\,km\,s$^{-1}$ respectively. Note that, for these models as well, the
initial total angular momentum is similar to the one contained in solar 
metallicity with rotational velocities of 300\,km\,s$^{-1}$. Since this is the
case, velocities between 600 and 800\,km\,s$^{-1}$ are considered in
this work as the average rotational velocities at these very low metallicities.

\subsection{Simulations}
The evolution of the models was followed until core Si--burning. Note that the
non--rotating and fast rotating 20 $M_\odot$ models at Z=10$^{-8}$ have a
strongly degenerate core such that Si--burning does not occur in the
central percent of the star. It is planned to improve the model to better
follow strongly degenerate core. This does nevertheless not affect the results
for presented in this work.
The 60 $M_\odot$ model was evolved until Ne--burning. This means that, for
this model only, the yields of the heavy elements can still
vary. The stellar yields are calculated as in Hirschi et al (2005).
Therefore the contribution from explosive nucleosynthesis is not included
and the remnant mass is determined from the CO core mass 
(see Maeder 1992). The impact on
the results are discussed in Hirschi (2004) and we only present yields for
elements which are not significantly affected by the evolution beyond 
our calculations. The main characteristics of the models are presented 
in Table \ref{table1}.
\begin{table}
\caption{Initial parameters of the models (columns 1--5): 
mass, metallicity, rotation velocity [km\,s$^{-1}$], 
total angular momentum [10$^{53}$\,erg\,s] and $\upsilon_{\rm ini}/ \upsilon_{\rm crit}$.
Total lifetime [Myr] and
various masses [$M_\odot$] (7--10): final mass (3), masses of the helium
and carbon--oxygen cores and the remnant mass.}
\begin{center}
\begin{tabular}{r r r r r r r r r r }
\hline \hline 
$M_{\rm{ini}}$ & $Z_{\rm{ini}}$ & $\upsilon_{\rm{ini}}$ & $J_{\rm{tot}}^{\rm{ini}}$ & $\upsilon_{\rm ini}/ \upsilon_{\rm crit}$
& $\tau_{\rm{life}}$ 
& $M_{\rm{final}}$ & $M_{\alpha}$ & $M_{\rm{CO}}$ & $M_{\rm{rem}}$  \\ 
\hline
20 & 2e-02 & 300 & 0.36 & 0.44 & 11.0 &  8.7626 & 8.66 & 6.59 & 2.57 \\
20 & 1e-03 & 000 &  --  & 0.00 & 10.0 & 19.5567 & 6.58 & 4.39 & 2.01 \\
20 & 1e-03 & 300 & 0.34 & 0.39 & 11.5 & 17.1900 & 8.32 & 6.24 & 2.48 \\
20 & 1e-05 & 000 &  --  & 0.00 & 9.80 & 19.9795 & 6.24 & 4.28 & 1.98 \\
20 & 1e-05 & 300 & 0.27 & 0.34 & 11.1 & 19.9297 & 7.90 & 5.68 & 2.34 \\
20 & 1e-05 & 500 & 0.42 & 0.57 & 11.6 & 19.5749 & 7.85 & 5.91 & 2.39 \\
20 & 1e-08 & 000 &  --  & 0.00 & 8.96 & 19.9994 & 4.43 & 4.05 & 1.92 \\
20 & 1e-08 & 300 & 0.18 & 0.28 & 9.98 & 19.9992 & 6.17 & 5.18 & 2.21 \\
20 & 1e-08 & 600 & 0.33 & 0.55 & 10.6 & 19.9521 & 4.83 & 4.36 & 2.00 \\
40 & 1e-08 & 700 & 1.15 & 0.55 & 5.77 & 35.7954 & 13.5 & 12.8 & 4.04 \\
60 & 1e-08 & 800 & 2.41 & 0.57 & 4.55 & 48.9747 & 25.6 & 24.0 & 7.38 \\
85 & 1e-08 & 800 & 4.15 & 0.53 & 3.86 & 19.8677 & 19.9 & 18.8 & 5.79 \\
\hline
\end{tabular}
\end{center}
\label{table1}
\end{table}

\subsection{20 $M_\odot$ models}
\subsubsection{Evolution of the internal structure}
\begin{figure}
  \centering
\resizebox*{0.5\textwidth}{!}{\includegraphics{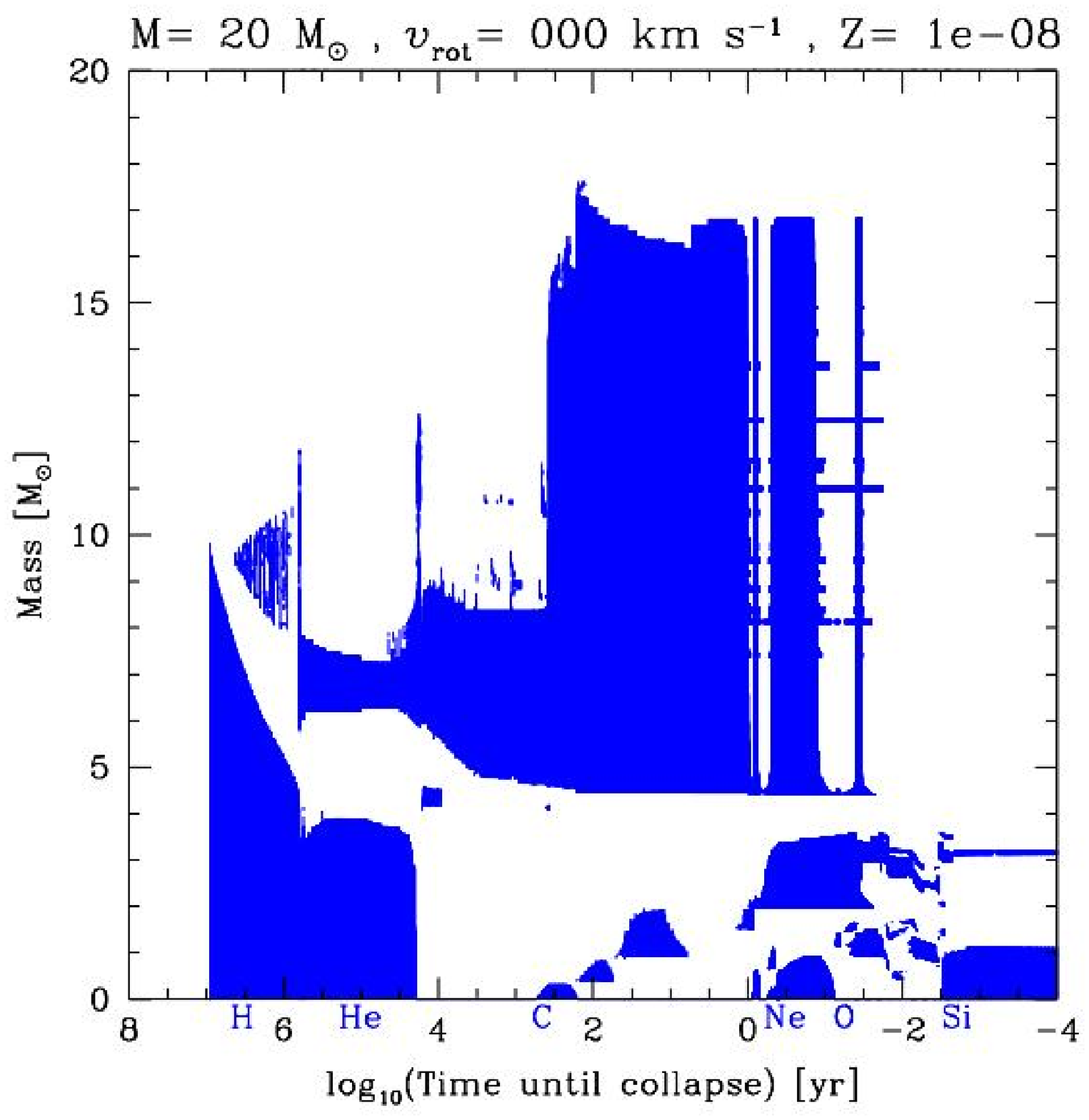}}\resizebox*{0.5\textwidth}{!}{\includegraphics{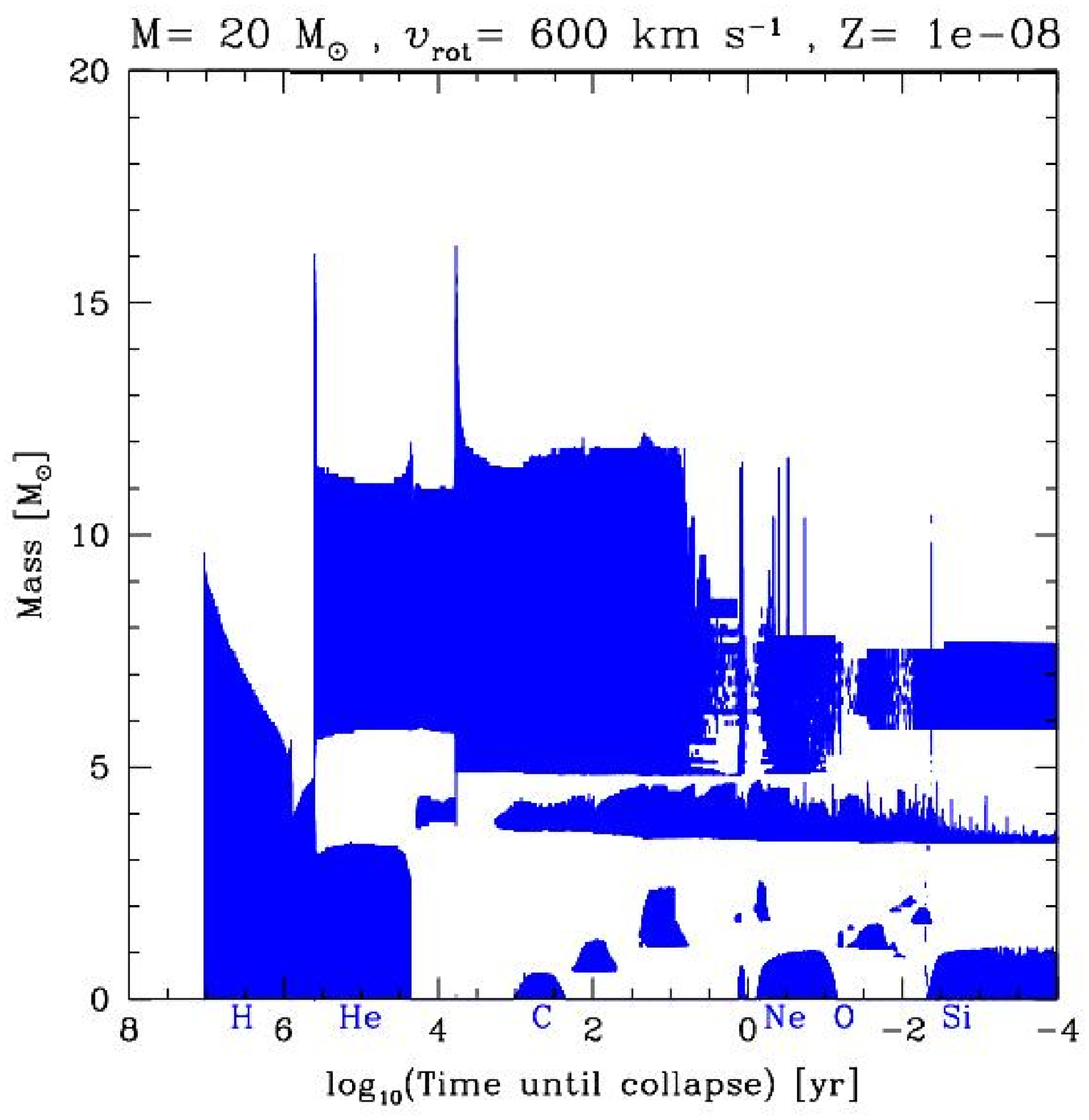}}
  \caption{Kippenhahn diagrams of 20$\,M_\odot$ models 
at $Z=10^{-8}$ with $\upsilon_{\rm ini}=$
 0\,km\,s$^{-1}$ ({\it left}) and
600\,km\,s$^{-1}$ ({\it right})  (figure taken from Hirschi 2006).}
\label{kip20}
\end{figure}
Mass loss becomes gradually unimportant as the metallicity decreases in
the 20 $M_\odot$ models. At solar metallicity, the rotating 20 $M_\odot$
model loses more than half of its mass, at $Z=0.001$, the models lose
less than 15\% of their mass, at $Z=10^{-5}$ less than 3\% and at 
$Z=10^{-8}$ less than 0.3\% (see Table \ref{table1}). This means that 
at very low metallicities,
the dominant effect of rotation is mixing for the mass range around 20
$M_\odot$. At solar metallicity and metallicities higher than about
$Z=10^{-5}$, rotational mixing increases the helium and CO core sizes
(see Table \ref{table1}). In particular, the oxygen yield is increased.

The impact of mixing on models at $Z=10^{-8}$ (and at $Z=0$ see 
Ekstr\"om et al 2006) is however different for an average rotation 
($\upsilon_{\rm ini}=$600\,km\,s$^{-1}$). The impact of mixing is best
pictured in the Kippenhahn diagram for this model (see Fig. \ref{kip20}
{\it right}). During hydrogen burning and the start of helium burning, mixing
increases the core sizes. Mixing of helium above the core suppresses
the intermediate convective zones linked to shell H--burning. So far the
process is similar to higher metallicity models. However, after some
time in He--burning, the mixing of primary carbon and oxygen into the
H--burning shell is important enough to boost significantly the strength
of the shell. As a result, the core expands and the size of the helium
burning core becomes and remains smaller than in the non--rotating
model (Fig. \ref{kip20} {\it left}). The yield of $^{16}$O being closely
correlated with the size of the CO core, it is therefore  
reduced due to the strong mixing. At the same time the carbon yield is increased.
This produces an upturn of C/O at very low metallicities.

\subsubsection{Stellar yields of CNO elements}

\begin{table}
\caption{Initial mass (column 1), metallicity (2) and rotation velocity 
[km\,s$^{-1}$] (3) and 
total stellar yields (wind + SN) [$M_\odot$] for carbon (4), nitrogen 
(5) and oxygen (6).}
\begin{center}
\begin{tabular}{r r r r r r r r r r }
\hline \hline 
$M_{\rm{ini}}$ & $Z_{\rm{ini}}$ & $\upsilon_{\rm{ini}}$ 
& $^{12}$C & $^{14}$N & $^{16}$O  \\ 
\hline
20 & 2e-02 & 300 & 4.33e-01 & 4.33e-02 & 2.57e+00 \\
20 & 1e-03 & 000 & 3.73e-01 & 3.31e-03 & 1.46e+00 \\
20 & 1e-03 & 300 & 6.76e-01 & 3.10e-03 & 2.70e+00 \\
20 & 1e-05 & 000 & 3.70e-01 & 4.27e-05 & 1.50e+00 \\
20 & 1e-05 & 300 & 4.81e-01 & 1.51e-04 & 2.37e+00 \\
20 & 1e-05 & 500 & 6.48e-01 & 5.31e-04 & 2.59e+00 \\
20 & 1e-08 & 000 & 2.62e-01 & 8.52e-03 & 1.20e+00 \\
20 & 1e-08 & 300 & 3.81e-01 & 1.20e-04 & 1.96e+00 \\
20 & 1e-08 & 600 & 8.23e-01 & 5.90e-02 & 1.35e+00 \\
40 & 1e-08 & 700 & 1.79e+00 & 1.87e-01 & 5.94e+00 \\
60 & 1e-08 & 800 & 3.58e+00 & 4.14e-02 & 1.28e+01 \\
85 & 1e-08 & 800 & 7.89e+00 & 1.75e+00 & 1.23e+01 \\
\hline
\end{tabular}
\end{center}
\label{yields}
\end{table}
\begin{figure}
  \centering
\resizebox*{1.0\textwidth}{!}{\includegraphics{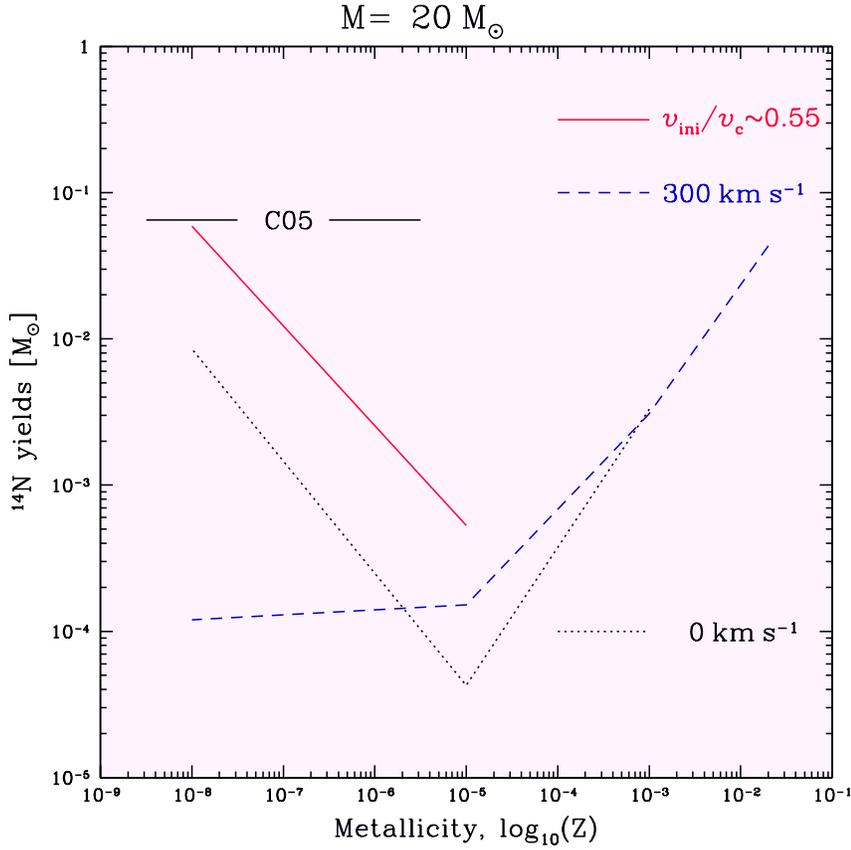}}
  \caption{Stellar yields of $^{14}$N 
as a function of the initial metallicity of the models (figure taken from Hirschi 2006).
The solid red, dashed blue and dotted black lines represent 
respectively
the models with an average rotation ($\upsilon_{\rm ini}/\upsilon_c \sim 0.55$ 
and $\upsilon_{\rm ini}$=500\,km\,s$^{-1}$ at $Z=10^{-5}$ 
and $\upsilon_{\rm ini}$=600\,km\,s$^{-1}$ at $Z=10^{-8}$),
with $\upsilon_{\rm ini}$=300\,km\,s$^{-1}$
and without rotation. The horizontal mark with C05 in the middle
corresponds to the primary $^{14}$N production needed in the chemical evolution models of
Chiappini et al (2005) to reproduce observations.
}
\label{yn}
\end{figure}
\begin{figure}
  \centering
\resizebox*{0.5\textwidth}{!}{\includegraphics{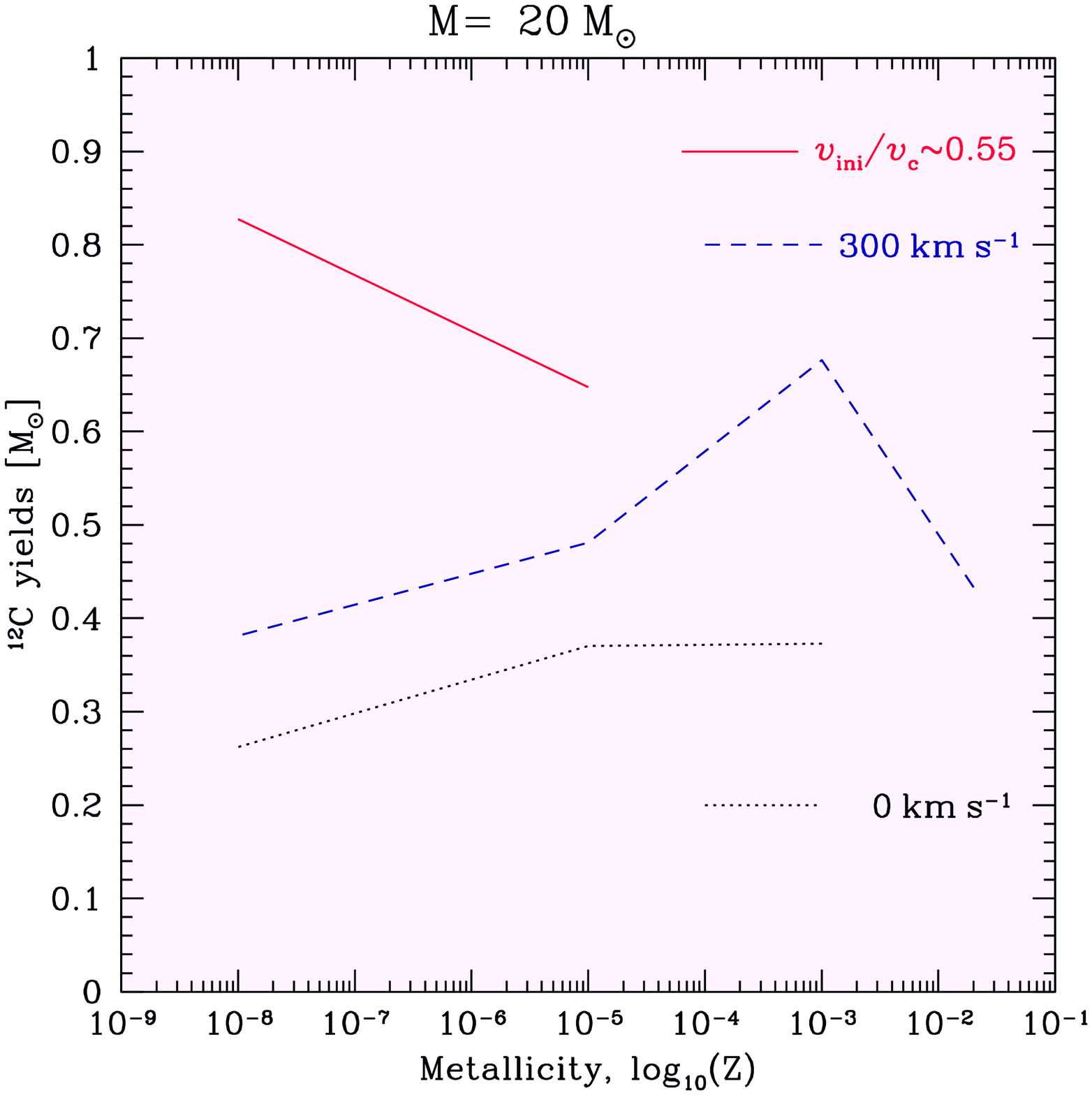}}\resizebox*{0.5\textwidth}{!}{\includegraphics{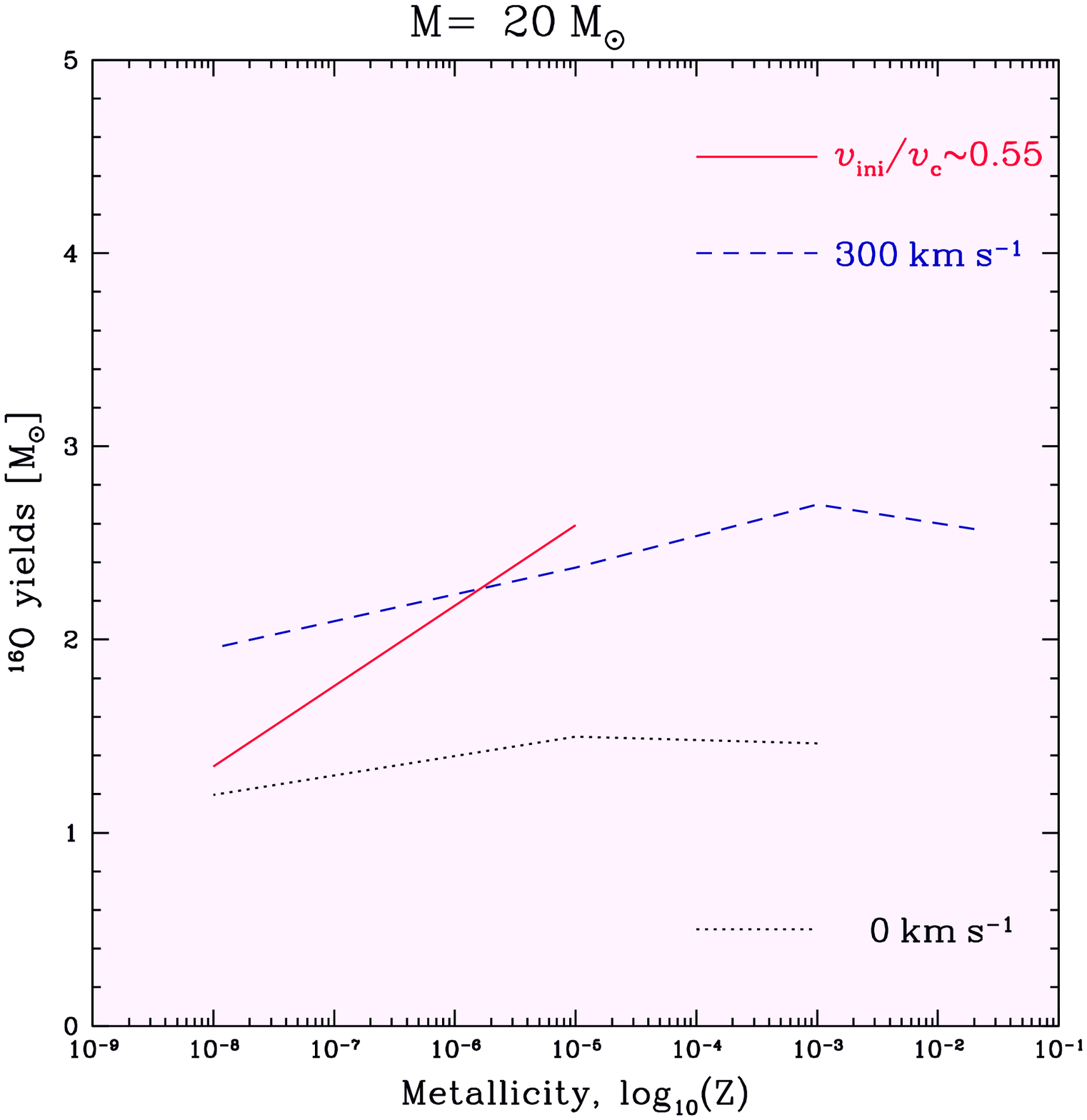}}
  \caption{Stellar yields of $^{12}$C ({\it left}) 
and $^{16}$O ({\it right})
as a function of the initial metallicity of the models.
The solid red, dashed blue and dotted black lines represent respectively
the models with an average rotation ($\upsilon_{\rm ini}/\upsilon_c \sim 0.55$ and
$\upsilon_{\rm ini}$=500\,km\,s$^{-1}$ at $Z=10^{-5}$ 
and $\upsilon_{\rm ini}$=600\,km\,s$^{-1}$ at $Z=10^{-8}$),
with $\upsilon_{\rm ini}$=300\,km\,s$^{-1}$
and without rotation (figures taken from Hirschi 2006).}
\label{yco}
\end{figure}
The yields of $^{12}$C, $^{14}$N and $^{16}$O are presented in Figs. 
\ref{yn} and \ref{yco} and their numerical values are given in Table
\ref{yields}
(see Hirschi 2006 for more details).
The most stringent observational constraint at very low Z is a very
high primary $^{14}$N production (Chiappini et al 2005, Prantzos 2004).
This requires extremely high primary $^{14}$N production in massive
stars, of the order of 0.06 $M_\odot$ per star. 
In Fig. \ref{yn}, we can see that only the model at $Z=10^{-8}$ and with
$\upsilon_{\rm ini}$=600\,km\,s$^{-1}$ can reach such high values.
The bulk of $^{14}$N is produced in the convective 
zone created by shell hydrogen burning (see Fig. 1 
{\it right}). If this 
convective zone deepens enough
to engulf carbon (and oxygen) rich layers, then significant amounts of 
primary
$^{14}$N can be produced ($\sim$0.01$\,M_\odot$). 
This occurs in both the non--rotating model 
and the fast rotating model but for different reasons.
In the non--rotating model, it occurs due to
structure rearrangements similar to the third dredge--up at the end of 
carbon burning. In the model with $\upsilon_{\rm ini}=$600\,km\,s$^{-1}$ 
it occurs during shell helium burning 
because of the strong mixing of carbon and oxygen into the 
hydrogen shell burning zone.

Models with higher initial masses at $Z=10^{-8}$ also produce large
quantities of primary nitrogen. More computations are necessary to see
over which metallicity range the large primary production takes place and
to see whether the scatter in the yields of the models with different masses
and metallicities is compatible with the observed scatter.

\subsection{Models at $Z=10^{-8}$}
\subsubsection{Stellar winds}
\begin{figure}
   \centering
  \resizebox*{0.5\textwidth}{!}{\includegraphics{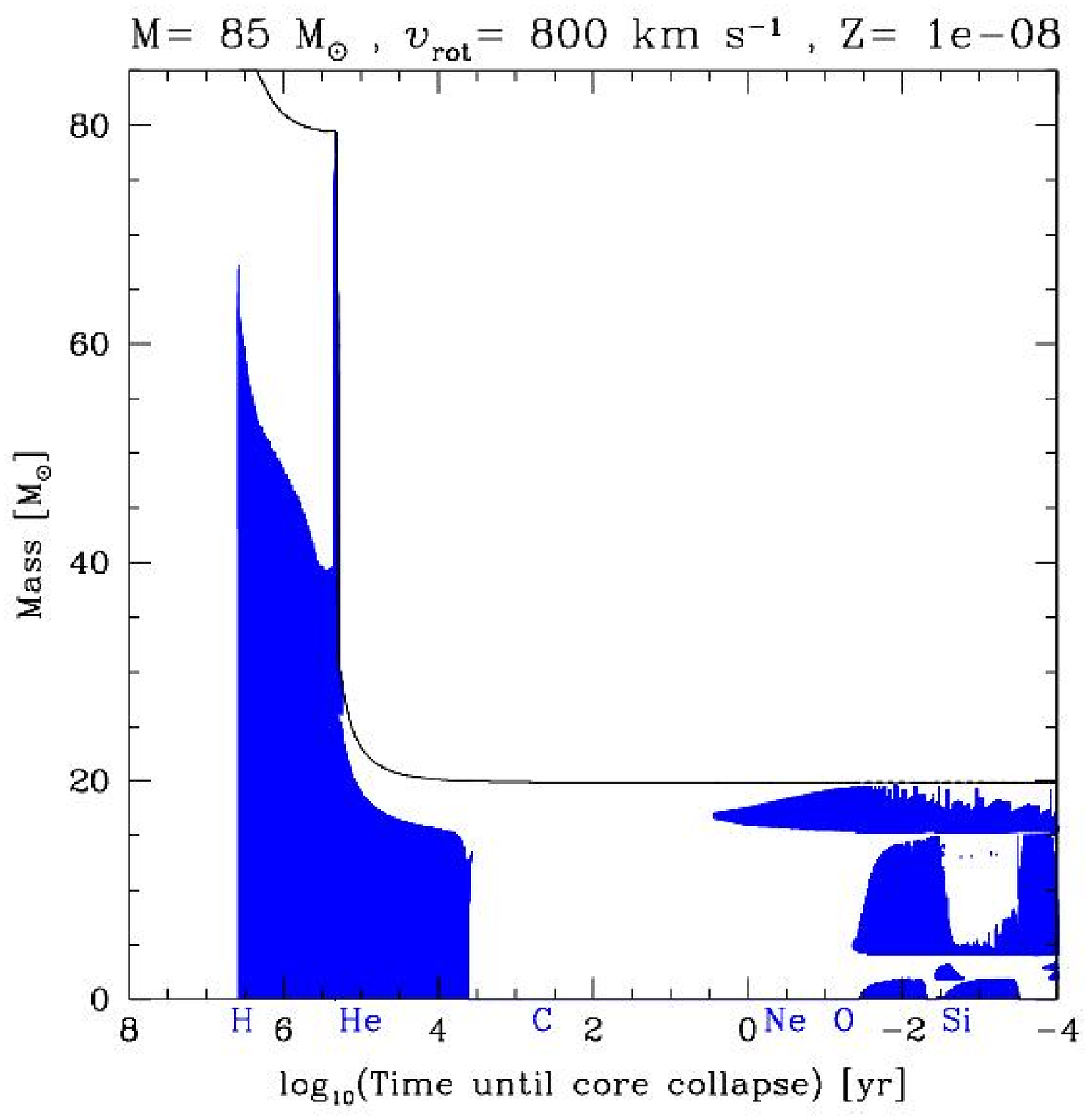}}\resizebox*{
0.5\textwidth}{!}{\includegraphics{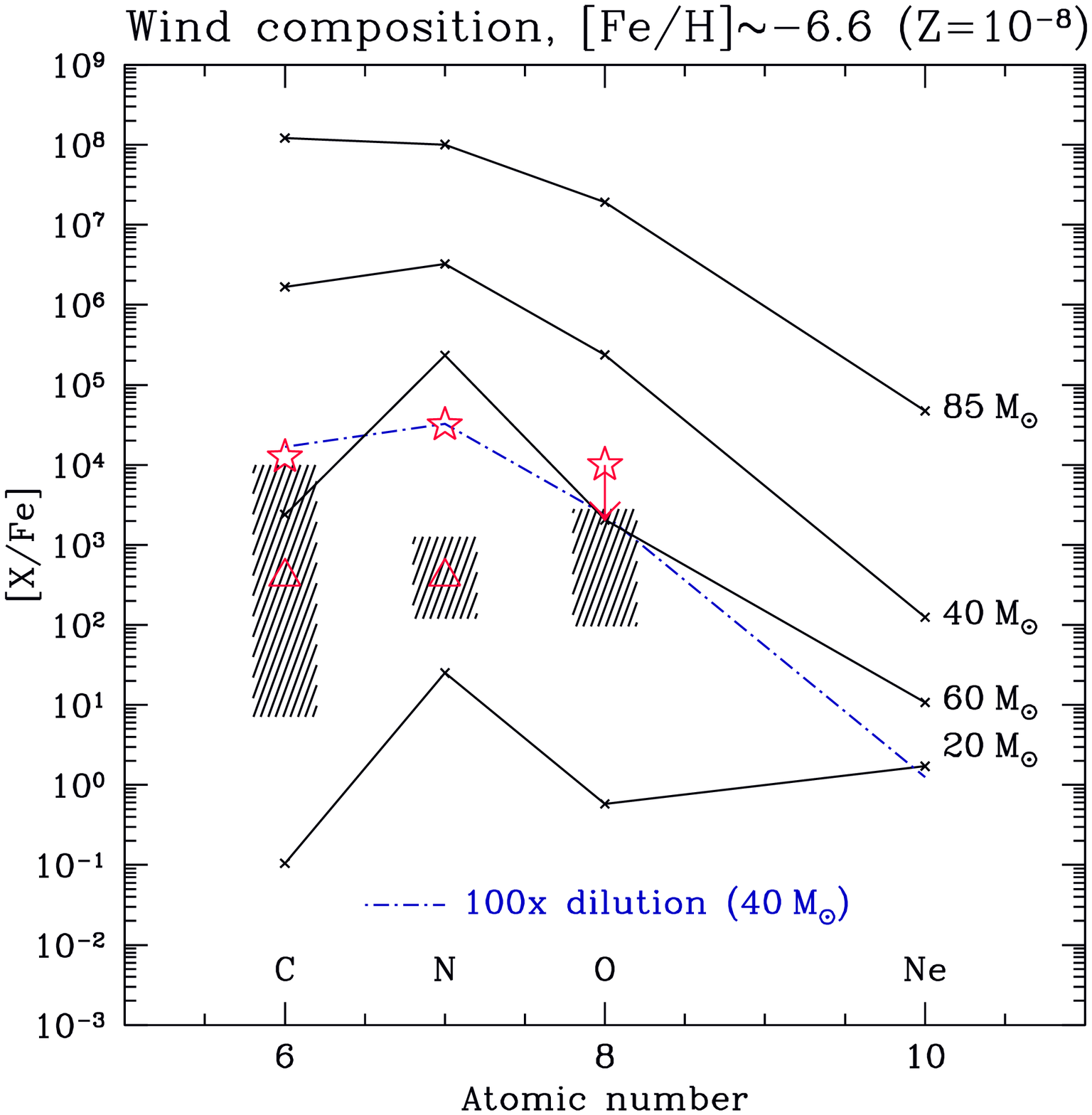}}
  \caption{{\it Left}: Kippenhahn diagram of 85$\,M_\odot$ models 
at $Z=10^{-8}$ with $\upsilon_{\rm ini}=$
 800\,km\,s$^{-1}$. 
 {\it Right}: 
 The solid lines represent the chemical composition of
 the wind material of the of the different models at
 $Z=10^{-8}$.
The hatched areas correspond to the range of values
measured at the surface of giant CEMP stars: HE 0107-5240, [Fe/H]$\simeq$~-5.3 
(Christlieb 2004);
CS 22949-037, [Fe/H]$\simeq$~-4.0 (Norris et al 2001, Depagne et al 2002);
CS 29498-043, [Fe/H]$\simeq$~-3.5 (Aoki et al 2004). 
The empty triangles (Plez \& Cohen 2005), [Fe/H]$\simeq -4.0$ 
and stars (Frebel et al 2005), [Fe/H]$\simeq -5.4$, only an
upper limit is given for [O/Fe], corresponding to
non-evolved CEMP stars (figures taken from Hirschi 2006).
}
\label{m85}
\end{figure}
Contrarily to what was initially expected from very
low metallicity stars, mass loss can occur in massive
stars (Meynet, Ekstr\"om \& Maeder 2005). The mass loss occurs in two
phases. The first phase is when the star reaches
break--up velocities towards the end of the main sequence. Due to
this effect stars, even metal free ones, are expected
to lose about 10\% of their initial masses for an
average initial rotation. The fraction could be higher
if the initial rotation turns out to be higher. The
second phase in which large mass loss can occur is
during the RSG stage. Indeed, stars more massive than
about 60 $M_\odot$ at $Z=10^{-8}$ become RSG and
dredge--up CNO elements to the surface. This brings
the total metallicity of the surface to values within
an order of magnitude of solar and triggers large mass
loss. The final masses of the models are given in
Table \ref{table1}. The case of the 85 $M_\odot$ model
is extremely interesting (see Fig. \ref{m85} {\it
left}) since it loses more than three quarter of its
initial mass. It even becomes a WO star.
\subsubsection{Wind composition and CRUMPS stars}
\begin{table}
\caption{Initial mass (column 1), metallicity (2) and rotation velocity 
[km\,s$^{-1}$] (3) and 
stellar wind ejected masses [$M_\odot$] for carbon (4), nitrogen 
(5) and oxygen (6).}
\begin{center}
\begin{tabular}{r r r r r r r r r r }
\hline \hline 
$M_{\rm{ini}}$ & $Z_{\rm{ini}}$ & $\upsilon_{\rm{ini}}$ 
& $^{12}$C & $^{14}$N & $^{16}$O  \\ 
\hline
20 & 1e-08 & 600 & 3.44e-12 & 3.19e-10 & 6.69e-11 \\
40 & 1e-08 & 700 & 5.34e-03 & 3.63e-03 & 2.42e-03 \\
60 & 1e-08 & 800 & 1.80e-05 & 6.87e-04 & 5.49e-05 \\
85 & 1e-08 & 800 & 6.34e+00 & 1.75e+00 & 3.02e+00 \\
\hline
\end{tabular}
\end{center}
\label{wind}
\end{table}
In Fig. \ref{m85} ({\it right}), we compare the
chemical composition of the wind material with
abundances observed in non-evolved carbon rich
extremely and ultra (Frebel et al 2005) metal poor stars. The
ejected masses of the wind material are also given in
Table \ref{wind}. It is very interesting to see that
the wind material can reproduce the observed abundance
in two ways. Either, the wind material is richer than
necessary and dilution (by a factor 100 for example
for the 40 $M_\odot$ models and HE1327-2326) with the ISM is needed or the wind 
has the right enrichment (for example the 60 $M_\odot$
and HE1327-2326) and the low mass star could form from
pure wind material. The advantage of the pure wind
material is that it has a ratio $^{12}$C/$^{13}$C around
5 (Meynet, Ekstr\"om \& Maeder 2005) and it can explain Li depletion.
With or without dilution, the wind material
has the advantage that it brings the initial
metallicity of the low mass star above the critical
value for its formation (Bromm 2005).

\section{Core collapse and explosive nucleosynthesis}

\subsection{Collapse, bounce, and postbounce evolution}
\label{sec:collapse.bounce.pbe}
Supernovae of type II, Ib/c are triggered by the gravitational collapse
of the inner stellar core when it reaches the Chandrasekhar mass,
i.e. the maximum iron core mass supported by the dominant electron
pressure. Early spherically symmetric simulations (at that time adiabatic
or based on local neutrino emission approximations) (Van Riper \& Arnett 1978
and References therein) suggested the
separation of the collapsing material into an inner and an outer core.
The inner core fails only marginally to be pressure supported and
the sound speed stays faster than the fluid velocity at all times.
In this regime, the infall velocity increases linearly with the radius.
Because the matter density and sound speed decrease with the distance
from the center, there is a sonic point where the sound speed approaches
the infall velocity. Outside of the sonic point, the infall velocity
is larger than the sound speed and information about the collapsed
core reach the outer layers only in form of a rarefaction wave. 

The essentially polytropic behavior of the equation of state allows
a non-relativistic analytic investigation of collapse (Goldreich
\& Weber 1980,Yahil 1983) (Note that general relativistic effects
decrease the mass of the inner core by \(\sim 20\% \) (Liebend\"orfer
et al. 2001, Hix et al. 2003)). The analytic investigation shows
that the mass of the inner core is well approximated by\[
M_{ic}\simeq \left( \kappa /\kappa _{0}\right) ^{3/2}M_{0},\]
where \( M \) refers to the core mass and \( \kappa  \) to the coefficient
in the polytropic equation of state \( p=\kappa \rho ^{\gamma } \)
with \( \gamma =4/3 \). Values with index \( 0 \) belong to the
marginally stable stage immediately before collapse. With the coefficient
for the degenerate ultra-relativistic electron gas (Shapiro \& Teukolsky
1983)\[
\kappa =\frac{\hbar c}{4}\left( 3\pi ^{2}\right) ^{1/3}\left( \frac{Y_{e}}{m_{B}}\right) ^{4/3},\]
one readily reproduces the conclusion that the mass of the inner core,
\( M_{ic} \), evolves proportionally to the squared electron fraction,
\( Y^{2}_{e} \). The evolution of \( Y_{e} \) and the entropy during
collapse are determined by the interplay between three different physical
processes: (i) the transition of protons to neutrons by electron capture
and neutrino emission, \( e^{-}+p\rightarrow n+\nu _{e} \), (ii)
isoenergetic neutrino scattering off nucleons, \( \nu _{e}+\{p,n\}\rightarrow \nu '_{e}+\{p,n\} \),
and (iii) neutrino thermalization by neutrino scattering off electrons,
\( \nu _{e}+e^{-}\rightarrow \nu '_{e}+e^{-\prime } \). Each of these
processes is rather straightforward to describe in a gas of free nucleons
(Bruenn 1985).
In reality, however, the by far most abundant nuclear species in the
cold collapsing matter are neutron-rich heavy nuclei. While the described
weak interactions basically stay the same, they rather involve nucleons
bound in nuclei than free ones. This leads to the interesting expectation
that the nuclear structure is probed in many nuclei that are difficult
or impossible to explore under terrestrial conditions. See (Mart\'inez-Pinedo
et al. 2004) for a recent review.

(i) As the density, \( \rho  \), increases, the electron chemical
potential increases with \( \mu _{e}\propto \rho ^{1/3} \) (Bethe
1990). The electrons fill higher energy levels and electron captures
on free or bound protons become more frequent. As long as the
density is lower than \( 5\times 10^{10} \) g/cm\( ^{3} \), the
neutrinos escape freely and the deleptonization rate is determined
by the electron capture rate. The electron capture rate on free protons
is by two orders of magnitude larger than the one on nuclei. But the
nuclei are by orders of magnitude more abundant targets. Because of
a highly floating magnitude of the free proton fraction, it is difficult
to decide from first principles whether electron captures on free
protons or nuclei would dominate in core collapse: Due to the repulsive
positive charges of the protons, nuclei are most stable in a neutron-rich
configuration and the ratio of bound protons to neutrons is mainly
determined by intrinsic properties of the nuclei and the entropy.
Hence, small changes in the electron fraction or entropy of the fluid
element cause significant changes in the small abundance of unbound
free protons. A rising abundance of free protons is immediately followed
by increased electron captures, lowering the electron fraction, and
thus driving back the free proton abundance. The deleptonization caused
by this strong negative feedback sets an upper limit to the sustainable
electron fraction in the inner core, even in the absence of other open channels for
electron capture (Messer 2000, Liebend\"orfer et al. 2002).

However, improved electron capture rates on heavy nuclei (Langanke \&
Mart\'inez-Pinedo 2002) that overcome the idealized blocking of Gamov-Teller
transitions in the traditionally applied single-particle model allow the deleptonization
to occur always faster than with captures on free protons alone. In
fact, the most recent simulations with improved electron capture
rates demonstrate that electron capture on heavy nuclei is always
dominating over electron capture on free protons and that the deleptonization
proceeds to significantly lower electron fractions at the center of
the core than with the previous ``standard'' nuclear physics input,
resulting in a \( 20\% \) smaller inner core at bounce (Langanke et al.
2003, Hix et al. 2003, Marek et al. 2005).

(ii) The basic neutrino opacity in core collapse is provided by neutrino
scattering on nucleons. Depending on the distribution of the nucleons
in space and the wavelength of the neutrinos, various important coherence
effects can occur: Most important during collapse is the binding of
nucleons in nuclei with a density contrast of several orders of magnitude
to the surrounding nucleon gas. The cross section for coherent scattering
of low energy neutrinos on nuclei scales with \( A^{2} \) for nuclei
with atomic number \( A \). Coherent scatterings off nuclei therefore
easily dominate the scattering opacity of neutrinos on nucleons of
the surrounding neutron gas (Freedman 1974). A useful comparison
of inverse mean free paths at the important density \( \rho =10^{12} \)
g/cm\( ^{3} \) is given in (Bruenn \& Mezzacappa 1997). They
find \( \lambda _{\nu +n}/\lambda _{\nu +A}\sim 3\times 10^{-2} \),
\( \lambda _{\nu +e}/\lambda _{\nu +A}\sim 2.5\times 10^{-2} \),
\( \lambda _{\nu +He}/\lambda _{\nu +A}\sim 10^{-4} \), and \( \lambda _{\nu +p}/\lambda _{\nu +A}\sim 5\times 10^{-5} \). 

Further corrections are necessary: with an increasing ratio between
the Coulomb potential of the positively charged ions and their thermal
energy, the average separation between nuclei will more strongly peak
around the value of most efficient packing. The neutrino opacities
are then to be corrected by an ion-ion correlation function \( \left\langle S_{ion}\left( E_{\nu }\right) \right\rangle <1 \)
(Itoh 1975, 1979, Horowitz 1997). Its consideration in core collapse
simulations based on a representative nucleus lowers the trapped lepton
fraction at bounce by \( 0.015 \) and increases the central entropy
per baryon by \( 0.12k_{B} \) (Bruenn \& Mezzacappa 1997). It
is found that the sizable entropy increase is not only due to the
increased deleptonization, but also to the fact that the correlation
effect is most pronounced for the low energy neutrinos with long
wavelengths. As current core collapse models proceed toward the
inclusion of a full ensemble of nuclei,
it becomes rather non-trivial how to adequately determine correlation
effects in the ion mixture, see for example (Sawyer 2005) and
References therein.

The situation is even more complicated in the phase transition from
isolated nuclei to bulk nuclear matter where the nuclei or the holes
in between them are strongly deformed. Various pasta-like shapes may
be assumed. Correlation effects in this phase could also affect the
neutrino opacities (Horowitz et al. 2004, Watanabe et al. 2004).
For an immediate effect on core collapse, however, it would be required
that this transition phase would reach to fairly low densities in
order to affect the opacities at the neutrinospheres where the neutrino
luminosities and spectra are set. A detailed overview over neutrino
opacities in nuclear matter has recently been given by (Burrows, Reddy
\& Thompson 2004). An extensive quantitative overview over most reaction
rates has also been provided in the literature (Bruenn \& Haxton
1991).

(iii) Neutrinos are produced by the capture of degenerate electrons
from high energy levels. To some extent depending on the Q-value of
the capturing nucleus, the emitted neutrino starts with a high energy
of the order of the electron chemical potential. As the neutrino opacities
scale with the squared neutrino energy, the initially
trapped neutrinos will down-scatter to lower energies until the diffusion
time scale becomes comparable to the thermalization time scale. The
thermalization in current collapse models occurs through neutrino-electron
scattering because the energy transfer per collision with the light
electron is more efficient than with the heavier nucleons. The contribution
of inelastic scattering of neutrinos off heavy nuclei depends on the
individual nuclei and remains to be explored in detail.

\begin{figure}
\centering
\resizebox*{0.465\textwidth}{!}{\includegraphics{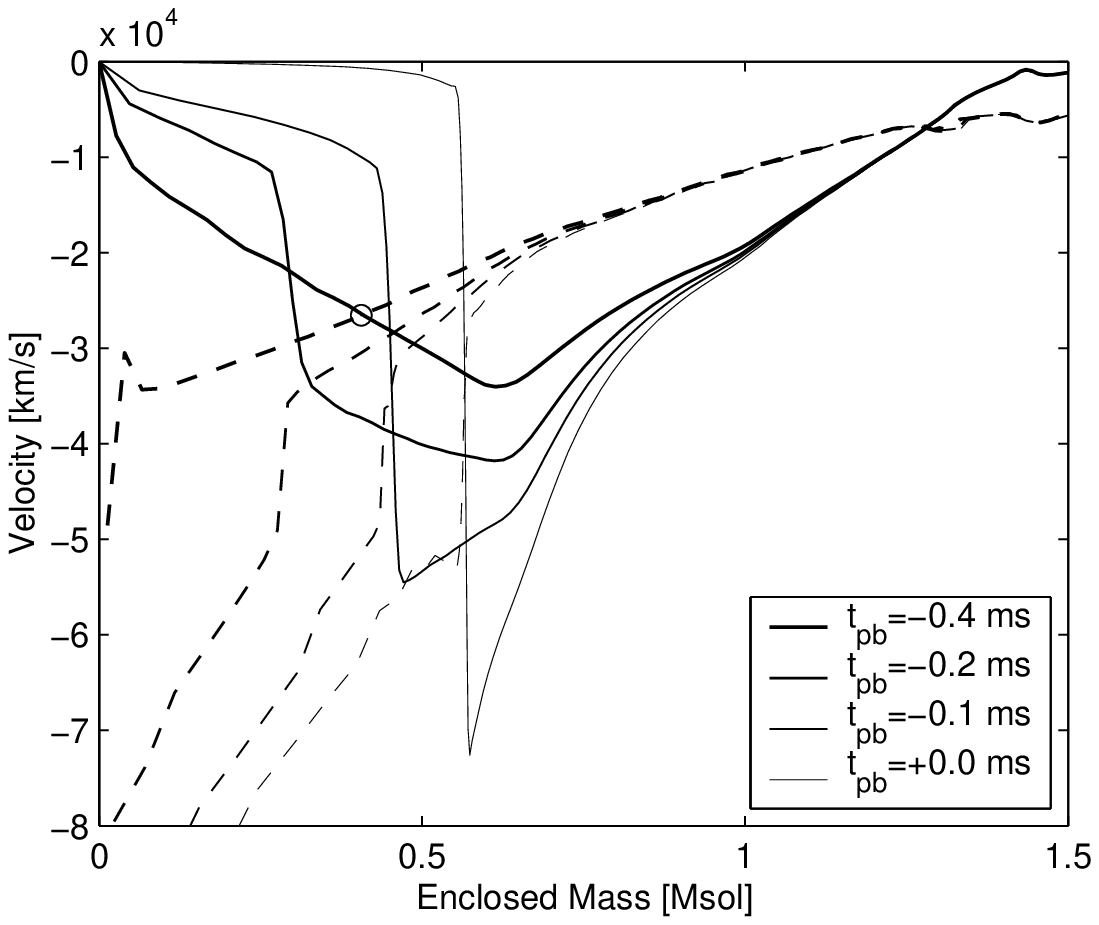}}
\resizebox*{0.5\textwidth}{!}{\includegraphics{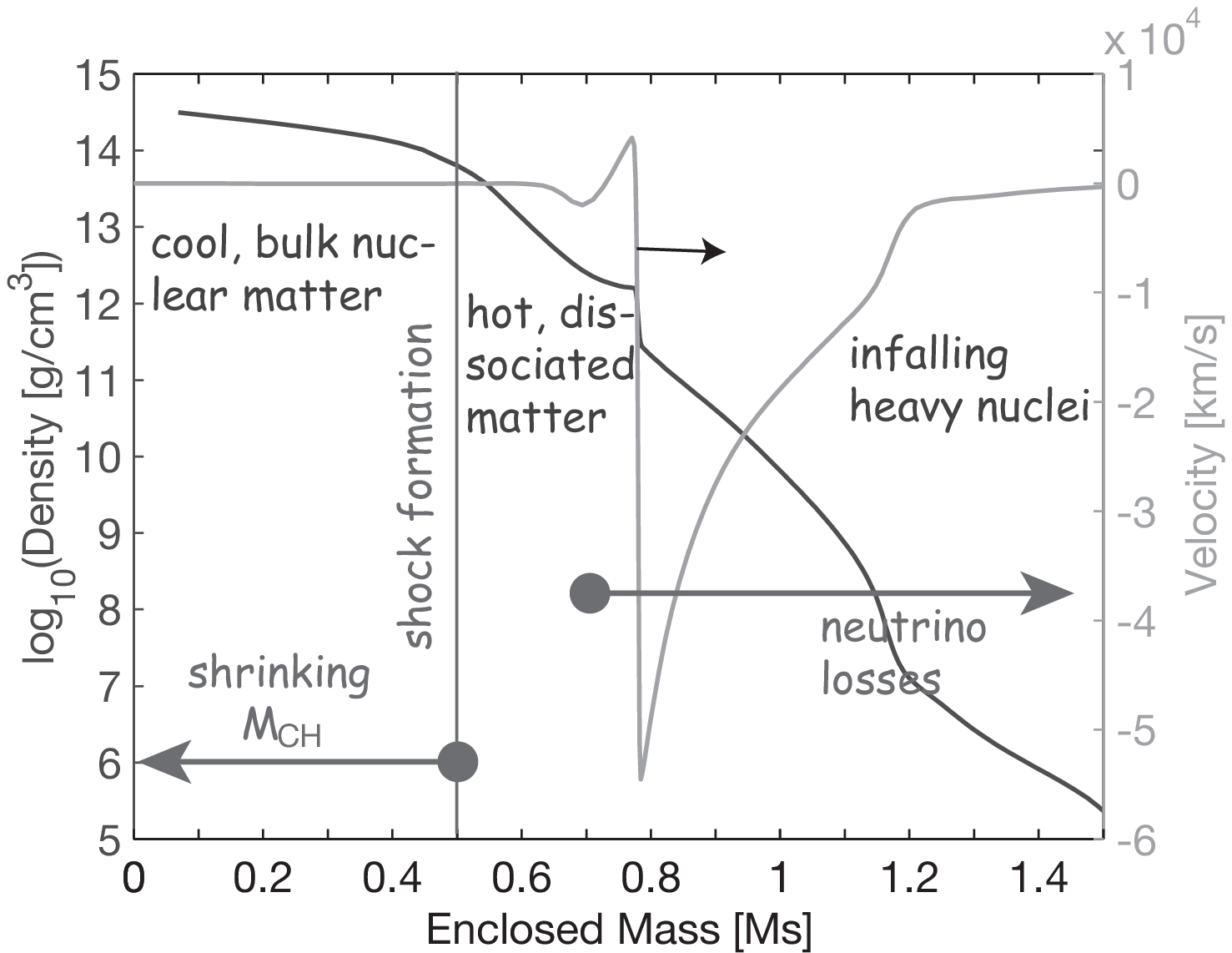}} 
\caption{The figure on the left hand side shows velocity profiles
immediately before bounce (solid lines).
The maximum infall velocity at \protect\( 0.4\protect \)
ms before bounce marks the edge of the inner core. A
circle is drawn at the sonic point, where the infall velocity is
equal to the sound speed (dashed lines). The later profiles demonstrate
how a pressure wave is launched at the center and runs through
the inner core until it turns into a shock front close to its edge.
The figure on the right hand side shows the velocity and density
profiles few ms after bounce when the neutrino burst is
launched. At this time the remaining shock energy (region with
positive velocities) is decaying and the shock stalls into an
expanding accretion front.}
\label{fig_velocity.ps}
\end{figure}

Once nuclear densities are reached at the center of the collapsing
core, repulsive nuclear forces dominate the stiffness of the equation
of state. The collapse is halted by an outgoing pressure wave. Different
snapshots of the velocity profile around bounce are shown in Fig. \ref{fig_velocity.ps}. The
pressure wave travels through the inner core, where infall velocities
are subsonic, and turns into a shock wave when it meets supersonic infall
velocities at its edge. The matter in the inner core experiences a
rather adiabatic pressure wave and remains at low entropy \( \sim 1.4 \)
kB per baryon. The shock wave in the outer core, however, heats matter
to entropies larger than \( \sim 6 \) kB per baryon so that heavy
nuclei are dissociated. If the bounce-shock were to dynamically propagate
through the core to expel outer layers in a prompt explosion, it would
have to provide the energy to dissociate the material between the
edges of the inner core and the iron core at a rate of \( 1.5\times 10^{51} \)
erg per \( 0.1 \) M\( _{\odot } \) of dissociated material (the
shock additionally suffers from unavoidable neutrino losses).

Numerical simulations based on standard input physics and accurate
neutrino transport exclude the possibility that the kinetic energy
of the hydrodynamical bounce at nuclear densities drives a prompt
supernova explosion (Liebend\"orfer et al. 2001, Bruenn et al. 2001,
Rampp \& Janka 2002, Thompson et al. 2003, Sumiyoshi et al. 2005).
While the trajectory through
core collapse determines the state of the cold nuclear matter inside
the protoneutron star (PNS), the mass of the hot mantle surrounding
the PNS grows by continued accretion. The infalling matter is heated
and dissociated by the impact at the fairly stationary accretion front.
Before about \( 50 \) ms after bounce, the entropy achieved by this
shock-heating is higher than the maximum entropy obtained by infinite
exposure to the prevailing neutrino field. With time ongoing, the
neutrinospheres recede to smaller radii and produce a harder neutrino
spectrum. On the other hand, the accretion front hydrostatically moves
to larger radii where the dissipated kinetic energy of the infalling
matter is smaller. Hence, only after \( 50 \) ms, when the entropy
of shocked material is smaller and the neutrino luminosity higher,
neutrino heating becomes effective behind the shock. Behind the shock,
matter continues to drift inward and to converge to higher densities.
The increasing electron chemical potential blocks the electron phase
space and favors antineutrino capture in the gain region below the
accretion front, and electron capture in the cooling region above
the PNS surface. Only expanding matter can lead to preferential neutrino
capture as will be discussed in section \ref{sec:SNIInucleosynth}.
The gain radius separates the
gain region from the cooling region. It is the radius where cooling
by neutrino and antineutrino emission balances heating by neutrino
and antineutrino absorption, i.e. the point where the infalling fluid
element reaches equilibrium entropy in the neutrino field produced at the
PNS surface. The reaction time scale in the cooling region is comparable
to or faster than the infall time scale so that an infalling fluid element
adheres to the equilibrium entropy on its further journey
toward the surface of the NS.

One half to two thirds of the neutrino luminosity in the heating region
stems from the accreting matter in the cooling region, the smaller
part diffuses out of the hot PNS. Because the accretion rate reacts
to neutrino heating and the neutrino luminosity to the accretion rate,
the conditions at the surface of the PNS are strongly coupled to the
evolution of the matter behind the accretion front. In spherical symmetry,
the important neutrino transport between these layers can be numerically
tackled in a fully consistent way by the accurate solution of the
Boltzmann transport equation for the three neutrino flavors (Rampp
\& Janka 2002, Thompson et al. 2003, Liebend\"orfer et al. 2004,
Sumiyoshi et al. 2005), where the latter two references implemented
all equations in general relativity. One result of these calculations
was that all progenitor stars between main sequence
masses of \( 9 \) and \( 40 \) M\( _{\odot } \) showed no explosions
in simulations of their postbounce evolution phases (see Fig.
\ref{fig:explosions}) (Liebend\"orfer et al. 2002). This indicates that
the neutrino flux emanating from the PNS has not the ample strength
to blow off the surrounding layers for a vigorous explosion without
the consideration of more details. Among the reasons are the strong
deleptonization during collapse, the dissociation of heavy nuclei
by the shock, a slow penetration of the neutrinos from layers at high
density through layers with densities between \( 10^{12}-10^{13} \)
g/cm\( ^{3} \), efficient neutrino cooling at the surface of the
PNS, and fast infall velocities through the gain region (see e.g.
Janka 2001).

However, spherically symmetric simulations ignore fluid instabilities
that are known to exist between the protoneutron star surface and
the stalled shock, as well as deep in the protoneutron star (Herant
et al. 1994). There
is consensus that the convective overturn in the heating region increases
the heating efficiency. Heated uprising plumes free the electron phase
space for neutrino absorptions and convect the energy toward the stalled
shock front instead of balancing the heating in loco by neutrino emission.
Additionally, cool and narrow downflows can feed the heating by
continued neutrino accretion luminosity. However, first simulations in axisymmetry
with energy-dependent neutrino transport do still not obtain vigorous
explosions (see Buras et al. 2005 and References therein for a discussion
of convection in the heating region). With respect to the
convection in the PNS, the luminosities are not significantly boosted
because the crucial region around the neutrinospheres appears to remain
convectively stable (see Bruenn et al. 2004 and references therein for
a discussion of PNS convection). As two-dimensional axisymmetric
calculations are now emerging as the new standard for simulations with
reliable neutrino transport, several newly appreciated phenomena are
currently explored in these computationally challenging calculations
(Buras et al. 2005, Walder et al. 2004, Swesty \& Myra 2005, Burrows et al. 2005).
The consistent coupling of the cooling and heating regions
with the feedback between accretion and neutrino luminosity is much
more involved in multiple dimensions than under the assumption of
spherical symmetry.

\subsection{Three-dimensional collapse simulations with magnetic fields}
\label{sec:3Dmhd}

Although axisymmetric supernova models add an essential new dimension
to the world of spherically symmetric models, namely the possibility
to accrete and expand matter at the same time at different locations,
there are still important degrees of freedom missing. Hot and cool
domains in the heating and cooling regions can only assume toroidal
shapes around the axis of symmetry so that convection will always
imply the motion of the whole torus. Axisymmetric simulations cannot
resolve small convective volumes or funnels linking the outer layers
with the surface of the PNS.

Not only the fluid
instabilities may bear an inherently three-dimensional topology, also
the presence of magnetic fields may introduce new degrees of freedom
that can only be explored in three dimensions. First estimations of magnetic
fields at the progenitor stage lead to rather low values around \( 10^9 \) Gauss
(Heger et al. 2005) and stars with magnetic fields are subject to magnetic
breaking of differential rotation (Maeder \& Meynet 2005). The uncertainties
are yet large, but nevertheless,
several explosion mechanisms have been suggested based on the
influence of magnetic fields: Large magnetic fields could collimate the matter
outflow after the postbounce evolution and lead to jet-like explosions
(LeBlanc \& Wilson 1970).
Smaller fields could grow by winding through differential rotation
and add magnetic pressure to the fluid pressure (Akiyama et al. 2003,
Ardeljan et al. 2004). In combination
with differential rotation the magnetic fields could lead to the magneto-rotational
instability so that the dissipation of the turbulent energy adds to
neutrino heating (Thompson et al. 2004). Moreover, the magnetic field
has been thought to possibly induce asymmetries to the neutrino heating
(Kotake et al. 2004). Finally, it has been suggested that magnetic loops
on the magnetized PNS would heat the hot mantle in analogy to the
solar corona (Ramirez-Ruiz \& Socrates. 2005). For all these reasons it
is interesting to
complement the axisymmetric simulations and accurate neutrino transport
with simulations that support all degrees of freedom of three-dimensional
space.

Already in one- and two-dimensional supernova models it becomes clear
that the computation time spent on hydrodynamics is negligible. Most
time is spent on energy-dependent neutrino transport. A systematic
improvement of the neutrino transport from one to two dimensions alone
required a substantial increase of computation time in yet incomplete
implementations (Buras et al. 2005, Livne et al. 2004) and three-dimensional
neutrino transport has not yet been attempted with a reliable resolution
of the neutrino phase space. Following three-dimensional approaches
of (Fryer \& Warren 2002, Scheck et al. 2003), we try to balance
the computation time spent on hydrodynamics and neutrino transport
by maximizing the degrees of freedom in the fluid dynamics in combination
with approximations in the neutrino transport. The effect of magnetic
fields on the dynamics of the nucleons in the hot mantle has not yet
been studied in three-dimensional numerical simulations with neutrino
transport approximations. Here we continue to report on efforts in
this direction (Liebend\"orfer, Pen \& Thompson 2005).

A simple and fast three-dimensional magneto-hydrodynamics code (Pen, Arras \& Wong 2003)
provides the core of our simulations. It spans a central region of
\( 600 \) km\(^3\) with an equidistant resolution of
\( 1 \) km in Cartesian coordinates. This covers the hot mantle and
part of the infalling layers. The code has received a parallelization
with MPI for cubic domain decomposition that minimizes the resources
occupied on distributed memory machines by a simple and efficient
reuse of buffer zones during the directional sweeps. The finite differencing
is second order accurate in time and space and handles discontinuities
in the conservation equations with a total variation diminishing scheme.
Furthermore, a specific choice of the finite differencing for the
update of the magnetic field conserves its divergence to machine precision.
The computational domain of the MHD code is embedded in spherically
symmetrically infalling outer layers evolved by a
one-dimensional hydrodynamics code.

\begin{figure}
\centering
\resizebox*{0.475\textwidth}{!}{\includegraphics{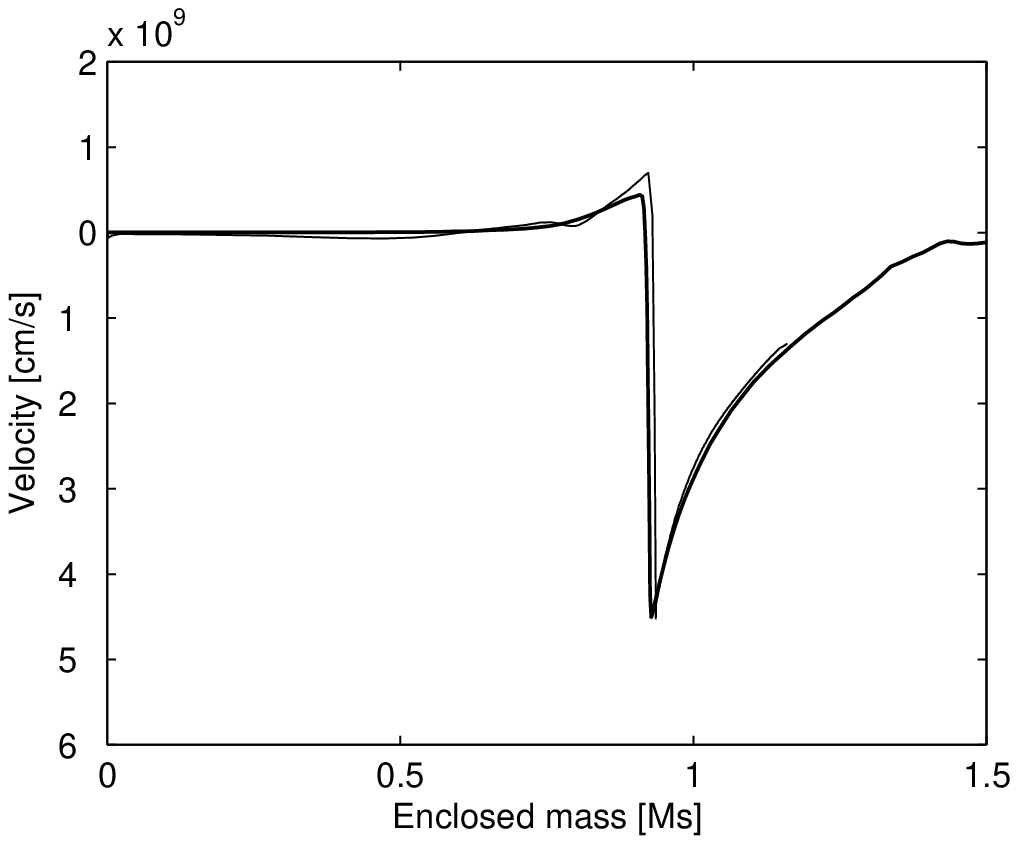}}
\resizebox*{0.49\textwidth}{!}{\includegraphics{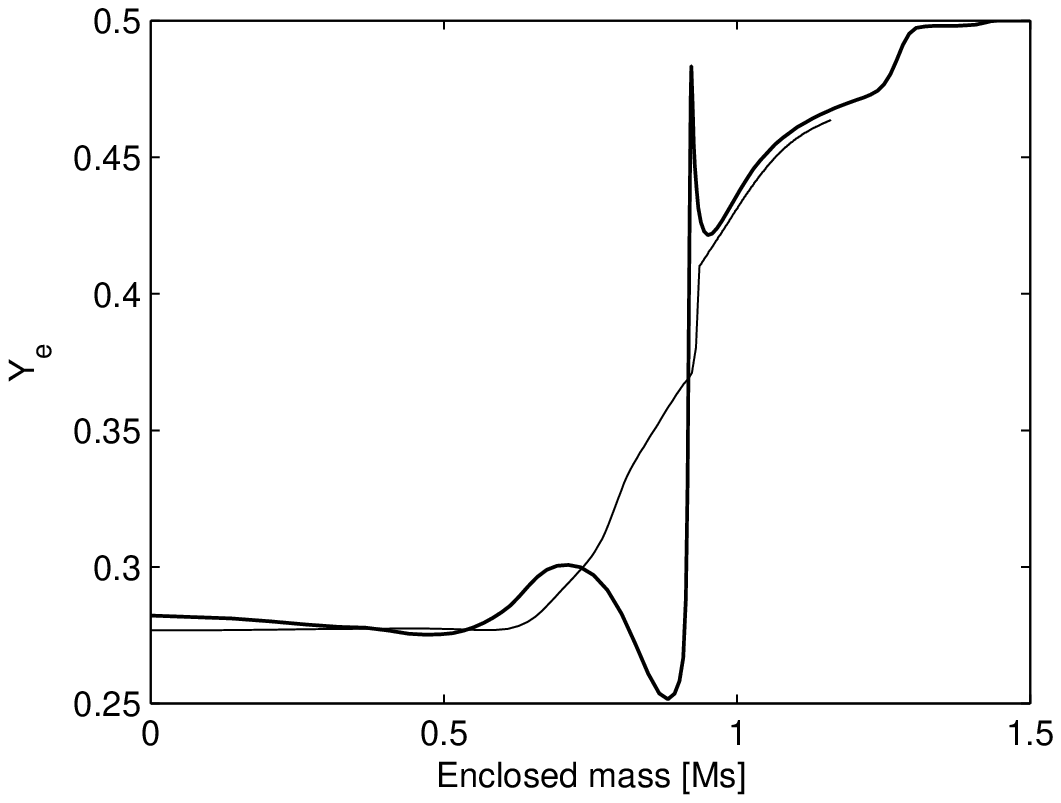}} 
\caption{Comparison of the three-dimensional simulation with
parameterized neutrino physics and effective gravitational
potential (thin lines) to the corresponding general relativistic spherically
symmetric simulation with Boltzmann neutrino transport (thick lines,
representing model G15 in Liebend\"orfer et al. 2005). The figure
on the left hand side shows the velocity profile at few ms after
bounce. The three-dimensional model reproduces very accurately
the shock position and shock strength of the model with detailed input physics.
The most prominent differences are shown in the figure on the right
hand side, where we compare the electron fraction. It can be seen
that, as expected, the deleptonization during collapse in the inner
core is well approximated. But the effects of the neutrino
burst are absent in the parameterized simulation. Electron captures
during the neutrino burst cause the strong \(Y_e\)-dip behind the shock
and neutrino absorptions cause the \( Y_e \)-peak just ahead of the shock.
See (Liebend\"orfer 2005) for a more detailed analysis of the parameterization.}
\label{fig:parameterization.ps}
\end{figure}

Our collapse simulations are launched from a \( 15 \) M\( _{\odot } \)
progenitor model (Woosley \& Weaver 1995). The Lattimer-Swesty
equation of state (Lattimer \& Swesty 1991) has been used. We imposed
a rotation with angular velocity \( \Omega =31.4 \) rad/s along
the z-axis with a quadratic cutoff at \( 100 \) km radius. Along
the same axis, we added a poloidal magnetic field of \( 10^{12} \)
Gauss. The deleptonization during collapse has been parameterized
in a very simple but carefully tested way (Liebend\"orfer 2005).
An investigation of the spherically symmetric model G15 (Liebend\"orfer et al. 2005)
with Boltzmann neutrino transport reveals that the electron fraction
during infall can roughly be approximated as a function of density
\( \rho  \). In our three-dimensional simulation, we update the electron
fraction with \( Y_{e}(x,y,z)=\min \left[ Y_{e}(x,y,z),Y_{e}^{G15}\left( \rho (x,y,z)\right) \right]  \),
where the function \( Y_{e}^{G15}(\rho ) \) has been read out of
model G15 at the time of core-bounce. Based on this parameterized
deleptonization (which also approximates the phenomenology of the various
nuclear and weak interaction processes discussed in Section \ref{sec:collapse.bounce.pbe}),
the changes of the entropy and the
momentum by neutrino stress are also estimated during the collapse
phase. The effective gravitational potential of (Marek et al. 2005)
has been used to implement general relativistic effects. Fig. \ref{fig:parameterization.ps}
compares the velocity and entropy profiles in a three-dimensional
collapse calculation without rotation and magnetic fields with the
corresponding general relativistic spherically symmetric calculation
based on Boltzmann neutrino transport. It becomes evident that the
scheme reproduces the collapse phase very accurately, while important
features of the postbounce phase are not yet captured, for example
the neutrino burst, or neutrino heating.

\begin{figure}
\centering
\resizebox*{\textwidth}{!}{\includegraphics{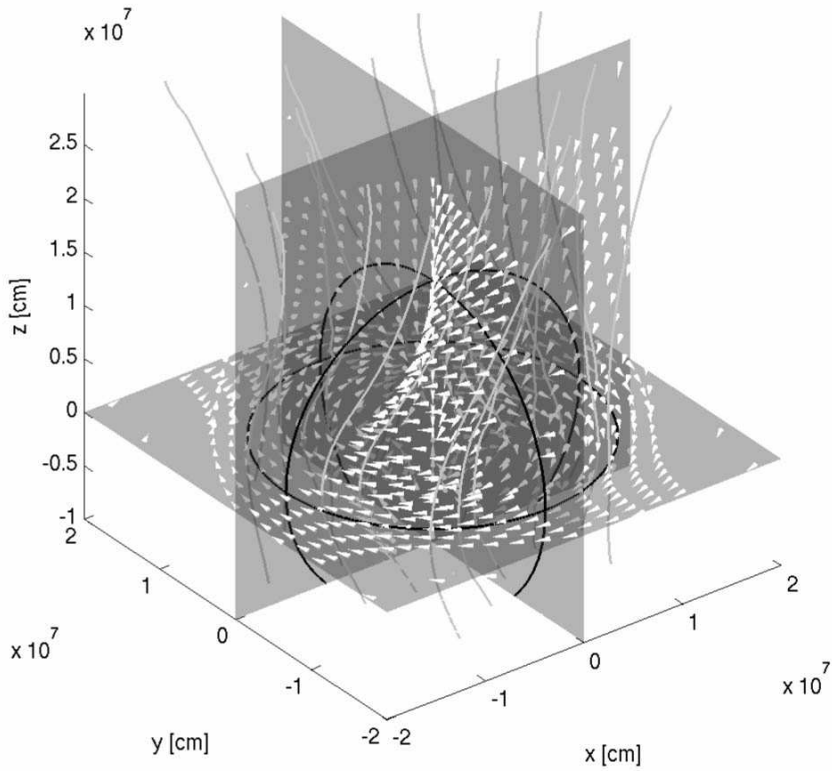}}
\caption{Snapshot of a three-dimensional simulation with rotation
and magnetic fields at \protect\( 150 \protect\) ms before bounce. The white arrows
act as mass tracers pointing out the direction of the velocities. At the
beginning of the simulation all mass tracers have been placed on the
three main planes. The vertical gray lines illustrate magnetic field lines
that are slowly distorted by the differential rotation. The black spheres
indicate the locations where the density is \protect\( 10^{10} \protect\) gcm\protect\( ^{-3} \protect\).}
\label{fig:collapse.ps}
\end{figure}
\begin{figure}
\centering
\resizebox*{\textwidth}{!}{\includegraphics{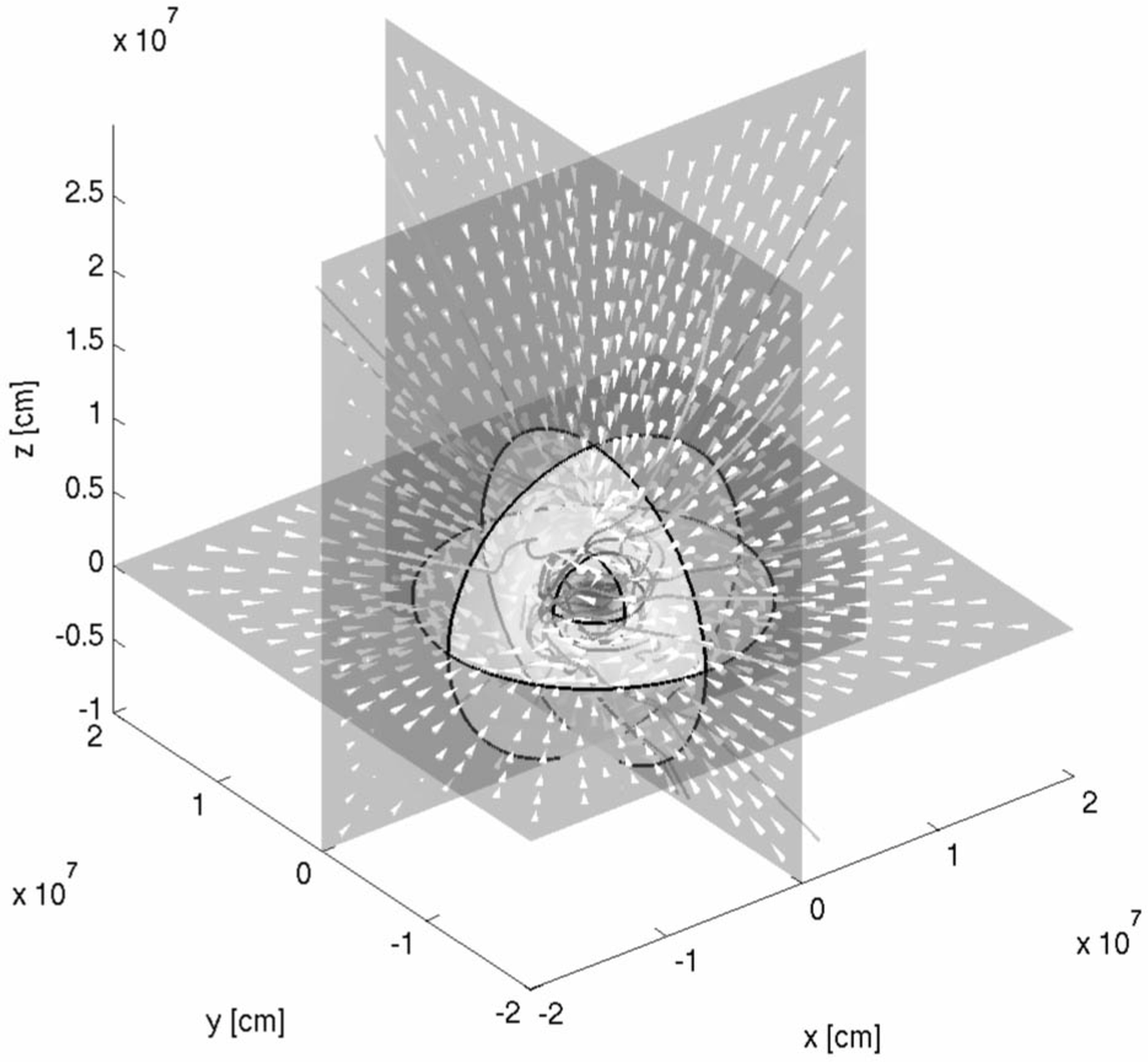}}
\caption{Snapshot of a three-dimensional simulation with rotation
and magnetic fields at \protect\( 5 \protect\) ms after bounce. The white arrows
act as mass tracers pointing out the direction of the velocities. During the
simulation mass tracers are inserted at the main planes. The black spheres
indicate isodensity contours at \protect\( 10^{10} \protect\) gcm\protect\( ^{-3} \protect\) and \protect\( 10^{12} \protect\) gcm\protect\( ^{-3} \protect\). The latter indicates the surface
of the protoneutron star at a smaller radius while the former sits in the
rapidly expanding accretion front. Note that only the accretion \protect\emph{front} is
expanding, not the hot stalled matter behind it. The shading of the main planes
indicates the matter entropy ranging from \protect\(\sim 1\protect\) (dark) to
\protect\(\sim 8 \protect\) (light) Boltzmann constants per baryon. The negative
entropy gradient leads to fluid instabilities that entangle the field lines in the
atmosphere of the PNS. The gray lines illustrate magnetic field lines wound
up by differential rotation around the surface of the protoneutron star.}
\label{fig:bounce.ps}
\end{figure}
Due to the rotation, the polar infall velocities are slightly larger
than the equatorial infall velocities. When the central density reaches
\( 10^{11} \) gcm\( ^{-3} \) the magnetic field lines become visibly
distorted. In the very early collapse phase this is due to matter
inflow along the rotation axis, meeting at the stellar center, and
pushing field lines outward into the equatorial plane. Moreover, due
to the centrifugal forces, the projection of the velocities onto the
plane orthogonal to the rotation axis is largest at about
\( 100 \) km above and below the gravitational center. These are
then the locations where the magnetic field lines condense most rapidly,
bending slightly outward around the center. With ongoing collapse,
this effect shifts to smaller radii and becomes more pronounced. At
bounce, the magnetic field exceeds \( 10^{15} \) Gauss in these hot
spots located \( \sim 10 \) km above and below the center. The field
lines run along double cones aligned with the z-axis, except for the
small deviation that circumvents the center. In the early shock expansion
until \( 5 \) ms after bounce, the shock front is almost spherically
symmetric. Afterward, the simulation becomes unrealistic, because
the dynamically important neutrino burst is not implemented. Behind
the expanding accretion front, entropy variations due to variations
in the shock strength induce fluid instabilities that entangle the
magnetic field lines. Figs. \ref{fig:collapse.ps} and \ref{fig:bounce.ps} show
a subdomain of the three-dimensional simulation at \( 150 \) ms before
bounce and at \( 5 \) ms after bounce, respectively.

The simulations have been performed on \( 64 \)
processors of the \( 528 \) processor McKenzie cluster at CITA. They
required a wall clock time of \( 288 \) hours. If the main computational
effort can be spent on magneto-hydrodynamics, three-dimensional
simulations allow a spatial resolution comparable to the resolution
used in spherically symmetric
simulations with Boltzmann neutrino transport. Of course, this is only
possible as long as the computationally
expensive neutrino transport can be replaced by adequate and efficient
parameterizations.

\subsection{Modeling Core Collapse Supernova Nucleosynthesis}
\label{sec:SNmodelling}

The complexity of neutrino transport and the frequent failure of
self-consistent models for core collapse supernovae to produce explosions have
generally divorced modeling of core collapse supernova nucleosynthesis from
modeling of the central engine. Despite the (fundamental) problem that the
supernova mechanism is still not understood, supernova nucleosynthesis
predictions have a long tradition. All of these predictions rely on
artificially introduced explosions, replacing the central engine either with a
parameterized kinetic energy piston 
(Woosley \& Weaver 1995, Rauscher et al.\ 2002, Chieffi \& Limongi 2004)
or a thermal bomb
(Thielemann et al.\ 1996, Nakamura et al.\ 2001). 
The explosion energy and the placement of the mass cut (separating ejected
matter from matter which is assumed to fall back onto the neutron star) are
tuned to recover the desired (observed) explosion energy and ejected $^{56}$Ni
mass. Both approaches are largely compatible (Aufderheide et al.\ 1991) 
and justifiable for the outer stellar regions. It is this inner region, where
most of the Fe-group nuclei are produced, which is most affected by the
details of the explosion mechanism, especially the effects of the interaction
of nuclei with the large neutrino flux.

The question arising is how more realistic supernova nucleosynthesis
predictions could be made, given the existing problem with self-consistent
explosions. Discussed improvements possibly leading to explosions are rotation
and magnetic fields (Thompson 2000, Thompson et al.\ 2004) or uncertainties in
neutrino opacities (see e.g.\ Burrows et al.\ 2004) or other
microphysics properties. They would introduce additional mixing at the
neutrino sphere and convective transport or change the neutrino luminosity via
improved opacities. This indicates two options for successful explosions:
(a) enhanced neutrino luminosities or
(b) enhanced deposition efficiencies for neutrino captures in the convective
layers.
We use two different methods to enforce explosions in otherwise non-explosive
models. In a first approach, we parameterize the neutral current neutrino
scattering opacities. This helps to artificially increase the diffusive fluxes
in regions of very high matter density, resulting in a faster deleptonization
of the proto-neutron star such that the neutrino luminosities are boosted in
the heating region. In a second approach, explosions are enforced by
multiplying the reaction rates for forward and backward reactions in 
$\nu_e +n \rightleftarrows p +e^-$ and
$\bar\nu_e + p \rightleftarrows n + e^+$ in the
heating region by equal factors. This reduces the time scale for neutrino
heating as it is expected in combination with overturn in this
convectively unstable domain. Both approaches allow successful explosions with
a consistently emerging mass cut without the need to add external energy.

\begin{figure}
  \centering
\resizebox*{1.0\textwidth}{!}{\includegraphics{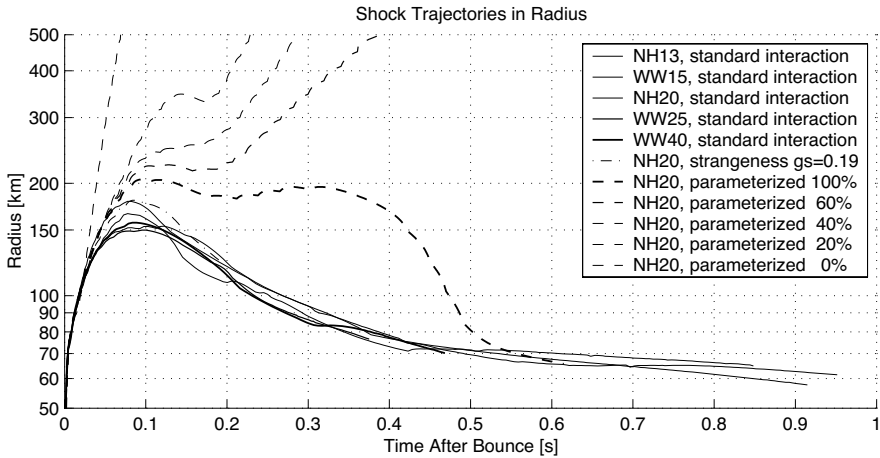}}
  \caption{A sequence of collapse calculations for different progenitor
  masses, showing in each case the radial position of the shock front
  after bounce as function of time. We see that the shocks are the strongest
  for the least massive stars. But in these 1D calculations all of them stall,
  recede, and turn into accretion shocks, i.e.\ not causing successful
  supernova explosions. A reduction in the neutrino-nucleon elastic
  scattering, leading to higher luminosities, can help explosions to
  occur.}
  \label{fig:explosions}
\end{figure}

Figure \ref{fig:explosions} shows the position of the shock front as function of
time for different core collapse supernova simulations. In all these models
the shock front runs outwards, stalls at $\sim 0.1$s, and subsequently
retrieves, leaving us with a non-explosive model. The dashed lines stem from
models with modified neutrino-nucleon elastic scattering as described
above. In the trajectories from these exploding models three phases can be
distinguished: an initial outwards propagating shock, which stalls around
0.1s after bounce. But shortly before 200ms after bounce, the accretion front
moves outwards again.

\subsection{Nucleosynthesis Implications of Neutrino Interactions}
\label{sec:SNIInucleosynth}

While the importance of neutrino interactions is manifest and well documented
in the $\nu$\emph{-process} and the \emph{r}-process, neutrinos potentially
impact all stages of supernova nucleosynthesis. Because of the impact of the
neutrinos on the nucleosynthesis, the nucleosynthesis products from future
explosion simulations (utilizing multi-group neutrino transport) will be
qualitatively different from either parametrized bomb or piston
nucleosynthesis models. The dominant process are $\nu$/$\overline{\nu}$ and
$e^{\pm}$ captures on shock dissociated free nucleons, though at later times
the more poorly known $\nu$/$\overline{\nu}$ captures on heavy nuclei may
contribute significantly.

An indispensable quantity to describe explosive nucleosynthesis in the
innermost ejecta is the electron fraction $Y_e$ (the number of electrons per
nucleon). This $Y_e$ is set by weak interactions in the explosively burning
layers, i.e.\ electron and positron capture, beta-decays, and neutrino or
anti-neutrino captures.

We examined the effects of both electron and neutrino captures in the context
of recent multi-group supernova simulations. These models are based on fully
general relativistic, spherically symmetric simulations (Liebend\"orfer et
al.\ 2001). Pruet et al.\ (2005) have
performed similar simulations using tracer particles from two dimensional
simulations (Buras et al.\ 2003). In both cases, artificial adjustments to the
simulations were needed to remedy the failure of the underlying models of
central engines to produce explosions (see also section \ref{sec:SNmodelling}
for details on how to invoke an explosion in a non-explosive model). Also in
both cases, the neutrino transport could not be run to later times and the
simulations were mapped to a more simple model at later times. Despite these
shortcomings, these simulations nevertheless reveal the significant impact of
neutrino interactions on the composition of the ejecta.

We find that all our simulations that lead to an explosion by neutrino heating
develop a proton-rich environment around the mass cut with $Y_e>0.5$
(Fr\"ohlich et al.\ 2006), as it is required from galactic evolution and
solar abundances. Several
phases can be identified in the evolution of the electron fraction of the
matter that will become the innermost ejecta. At early times matter is
degenerate and electron capture dominates. At the same time matter is being
heated by neutrino energy deposition and subsequently, the degeneracy is
lifted. While the ratio between electron captures and positron captures
significantly decreases, neutrino absorption reactions start to dominate the
change of $Y_e$. As the matter expands the density decreases and eventually the
electron chemical potential drops below half the mass difference between the
neutron and proton. With both neutrino absorption and emission processes
favoring a higher electron fraction, $Y_e$ rises markedly in this phase,
reaching values as high as 0.55.

\begin{figure}
  \centering
\resizebox*{0.5\textwidth}{!}{\includegraphics{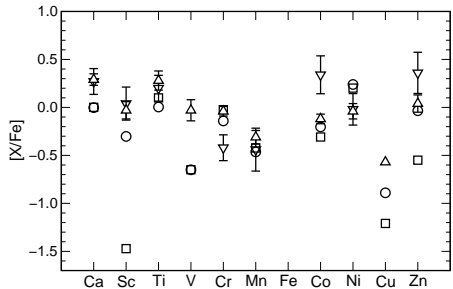}}
  \caption{Comparison of elemental overabundances in the mass range Ca to Zn
  for different calculations. The triangles with error bars represent
  observational data. The triangles facing upwards (Gratton \& Sneden 1991)
  originate from an analysis of stars with $-2.7<\mathrm{[Fe/H]}<-0.8$. The
  triangles facing downwards (Cayrel et al.\ 2004) is data for a sample of
  extremely metal poor stars ($-4.1< \mathrm{[Fe/H]}<-2.7$). The circles are
  abundances of our recent calculation (Fr\"ohlich et al.\ 2006). The squares
  are abundances of Thielemann et al.\ 1996.}
  \label{fig:elemental}
\end{figure}

The global effect of this proton-rich ejecta is the replacement of previously
documented overabundances of neutron rich iron peak nuclei (Woosley \& Weaver
1995, Thielemann et al.\ 1996). Production of
$^{58,62}$Ni is suppressed while $^{45}$Sc and ~$^{49}$Ti are enhanced. The
results (Fig.\ \ref{fig:elemental}) for the elemental abundances of scandium,
cobalt, copper, and zinc are closer to those observed (see
Fr\"ohlich et al. 2006 for details). We have found that the transformation of 
protons into
neutrons by neutrino captures allows (n,p)-reactions to take the place of the
$\beta$-decays of waiting point nuclei (with lifetimes longer than the
expansion timescale), allowing significant flow to $A>64$. We termed this
process the $\nu p$-process (see Fr\"ohlich et al.\ 2005 for more details) due to
the essential role of the neutrinos in producing these light p-nuclei. These
results clearly illustrate the need to include the full effect of the
supernova neutrino flux on the nucleosynthesis if we are to accurately
calculate the nucleosynthesis from core collapse supernovae.

\section{Conclusion}
Two series of pre-SN models and their yields were presented in this paper.
The  first series consists of 20 $M_\odot$ models with varying initial
metallicity (solar down to $Z=10^{-8}$) and rotation
($\upsilon_{ini}=0-600$\,km\,s$^{-1}$). The second one consists of
models with an initial metallicity of $Z=10^{-8}$, masses between 20 and
85 $M_\odot$ and average rotation velocities at these metallicities 
($\upsilon_{ini}=600-800$\,km\,s$^{-1}$).
The most interesting models are the models 
with $Z=10^{-8}$ ([Fe/H]$\sim-6.6$).
In the course of helium burning, carbon and oxygen are mixed into the
hydrogen burning shell. This boosts the importance of the shell and
causes a reduction of the size of the CO core. Later in the evolution,
the hydrogen shell deepens and produces large amount of primary
nitrogen. For the most massive models ($M\gtrsim 60$\,$M_\odot$),
significant mass
loss occurs during the red supergiant stage. This mass loss is due to
the surface enrichment in CNO elements via rotational and convective
mixing. 
The yields of the rotating 20 $M_\odot$ models can 
 reproduce the observed abundances at the surface of extremely 
metal poor (EMP) stars and the metallicity trends.
The wind of the massive models can also reproduce the CNO abundances of the
carbon--rich UMPs, in particular for the most metal-poor star known to
date, HE1327-2326.
The inclusion of neutrino interactions in explosive nucleosynthesis changes
significantly the final abundances by changing $Y_e$ to values above 0.5 in the
innermost ejecta. Neutrino interactions also enables nucleosynthesis 
to go beyond p--rich waiting point nuclei. The new results for iron group nuclei
are in much better agreement with observations of EMP stars 
(Sc, Ti, Ni, Zn, ...).

\subsection*{References}

{\small

\bref T.~{Abel}, G.~L. {Bryan}, and M.~L. {Norman}, \emph{Science} \textbf{295},
  93--98 (2002).

\bref S. Akiyama, J.~C. Wheeler, D.~L. Meier
  and I. Lichtenstadt, \emph{\apj} \textbf{584} 954 (2003).

\bref C.~{Angulo}, M.~{Arnould}, M.~{Rayet}, et al, \emph{Nuclear
  Physics A} \textbf{656}, 3--183 (1999).

\bref W.~{Aoki}, J.~E. {Norris}, S.~G. {Ryan}, et al, \emph{\apj} \textbf{608}, 971--977 (2004).

\bref W.~{Aoki}, A.~{Frebel}, N.~{Christlieb}, et al, \emph{astro-ph/0509206}  (2005).

\bref N.~V. Ardeljan, G.~S. Bisnovatyi-Kogan, S.~G. Moiseenko, \emph{MNRAS}
  \textbf{359} 333--344 (2005).
 
\bref Aufderheide, M.\,B., Baron, E, Thielemann, F.-K. 1991, ApJ 370, 630

\bref T.~C. {Beers}, G.~W. {Preston}, and S.~A. {Shectman}, \emph{\aj} \textbf{103},
  1987--2034 (1992).

\bref T.~C. {Beers}, ``Low-Metallicity and Horizontal-Branch Stars in the
  Halo of the Galaxy'', in \emph{ASP Conf. Ser. 165: The Third Stromlo
  Symposium: The Galactic Halo}, 1999, pp. 202--.

\bref H.~A. Bethe, \emph{Rev. Mod. Phys.} \textbf{62} 801 (1990).

\bref V.~{Bromm}, \emph{astro-ph/0509354}  (2005).

\bref S.~W. Bruenn, \emph{\apjs} \textbf{58} 771 (1985).

\bref S.~W. Bruenn, W.~C. Haxton, \emph{\apj} \textbf{376} 678 (1991).

\bref S.~W. {Bruenn}, A.~{Mezzacappa}, \emph{Phys. Rev. D} \textbf{56} 7529 (1997).

\bref S.~W. {Bruenn}, K.~R. {De Nisco}, A.~{Mezzacappa}, \emph{\apj} \textbf{560}
  326 (2001).

\bref S.~W. Bruenn, E.~A. Raley, A.~Mezzacappa, \emph{astro-ph/0404099}  (2004).

\bref Buras, R., Rampp, M., Janka, H.-T., Kifonidis, K. 2003, Phys.\ Rev.\ Lett.\ D
90, 241101

\bref R. Buras, M. Rampp, H.-Th. Janka, K. Kifonidis, \emph{astro-ph/0507135}, in press (2005).
 
\bref Burrows, A., Reddy, S., Thompson, T.\,A. 2004, Nucl.\ Phys.\ A, in press

\bref A. Burrows, E. Livne, L. Dessart, C. Ott, J. Murphy, \emph{astro-ph/0510687},
  in press    (2005).

\bref R.~{Cayrel}, E.~{Depagne}, M.~{Spite}, et al, A\&A  \textbf{416}, 1117--1138 (2004).

\bref C.~{Chiappini}, F.~{Matteucci}, and S.~K. {Ballero}, \emph{\aap} \textbf{437},
  429--436 (2005).

\bref Chieffi, A., Limongi, M. 2004, ApJ 608, 405

\bref N.~{Christlieb}, B.~{Gustafsson}, A.~J. {Korn}, et al, \emph{\apj}
  \textbf{603}, 708--728 (2004).

\bref E.~{Depagne}, V.~{Hill}, M.~{Spite}, et al, \emph{\aap} \textbf{390}, 187--198
  (2002).

\bref P.~{Fran{\c c}ois}, F.~{Matteucci}, R.~{Cayrel}, et al, \emph{\aap} \textbf{421}, 613--621 (2004).

\bref A.~{Frebel}, W.~{Aoki}, N.~{Christlieb}, et al, \emph{\nat}
  \textbf{434}, 871--873 (2005).

\bref D.~Z. {Freedman}, \emph{Phys. Rev. D} \textbf{9} 1389 (1974).

\bref C. Fr\"ohlich, P. Hauser, M. Liebend\"orfer,
G. Mart\'{\i}nez-Pinedo, F.-K. Thielemann, E. Bravo, N.~T. Zinner,
W.~R. Hix, K. Langanke, A. Mezzacappa, K. Nomoto, \emph{\apj}
\textbf{637} (2006)

\bref C. Fr\"ohlich, G. Mart\'{\i}nez-Pinedo, M. Liebend\"orfer,
F.-K. Thielemann, E. Bravo, W.~R. Hix, K. Langanke, N.~T. Zinner,
\emph{astro-ph/0511376} (2005)

\bref C.~F. Fryer and M.~S. Warren, \emph{\apj} \textbf{574} 65 (2002).

\bref I.~{Fukuda}, \emph{\pasp} \textbf{94}, 271--284 (1982).

\bref P.~{Goldreich}, S.~V. {Weber}, \emph{\apj} \textbf{238} 991 (1980).

\bref Gratton, R.\,G., Sneden, C. 1991, A\&A 241, 501

\bref A.~{Heger}, and S.~E. {Woosley}, \emph{\apj} \textbf{567}, 532--543 (2002).

\bref  A. Heger, S.~E. Woosley, H.~C. Spruit, \emph{\apj} 626 350--363 (2005).
 
\bref M.~{Herant}, W.~{Benz}, W.~R. {Hix}, C.~L. {Fryer}, S.~A. {Colgate}, \emph{\apj}
   \textbf{435} 339 (1994).

\bref R.~{Hirschi}, \emph{Ph.D.~Thesis, http://quasar.physik.unibas.ch/$\sim$hirschi/workd/thesis.pdf}  (2004).

\bref R.~{Hirschi}, G.~{Meynet}, and A.~{Maeder}, \emph{\aap} \textbf{425}, 649--670  (2004).

\bref R.~{Hirschi}, G.~{Meynet}, and A.~{Maeder}, \emph{\aap} \textbf{433},
  1013--1022 (2005).

\bref R.~{Hirschi}, \emph{\aap ,\ in prep}  (2006).

\bref W.~R. Hix, et~al., \emph{Phys. Rev. Lett.} \textbf{91} 201102 (2003).

\bref C.~J. {Horowitz}, \emph{Phys. Rev. D} \textbf{55} 4577 (1997).

\bref C.~J. {Horowitz}, M.~A. {P{\' e}rez-Garc{\'{\i}}a}, J.~{Piekarewicz}, \emph{Phys.
  Rev. C} \textbf{69} 045804 (2004).

\bref C.~A. {Iglesias}, and F.~J. {Rogers}, \emph{\apj} \textbf{464}, 943-- (1996).

\bref G.~{Israelian}, A.~{Ecuvillon}, R.~{Rebolo}, R.~{Garc{\'{\i}}a-L{\' o}pez},
  P.~{Bonifacio}, and P.~{Molaro}, \emph{\aap} \textbf{421}, 649--658 (2004).

\bref N.~{Itoh}, \emph{Progress of Theoretical Physics} \textbf{54} 1580 (1975).

\bref N.~{Itoh}, H.~{Totsuji}, S.~{Ichimaru}, H.~E. {Dewitt}, \emph{\apj}
  \textbf{234} 1079 (1979).

\bref Y.~I. {Izotov}, and T.~X. {Thuan}, \emph{\apj} \textbf{602}, 200--230 (2004).

\bref H.-Th. Janka, \emph{\aap} \textbf{368}, 527--560 (2001).

\bref  K. Kotake, S. Yamada, K. Sato, \emph{\apj} \textbf{618} 474--484 (2004).

\bref K.~Langanke, G.~Mart{\'\i}nez-Pinedo, \emph{Rev. Mod. Phys.} \textbf{75} 819 (2003).

\bref K.~Langanke, \emph{et~al.}, \emph{Phys. Rev. Lett.} \textbf{90} 241102 (2003).

\bref J. Lattimer and F.~D. Swesty, \emph{Nucl. Phys. A} \textbf{535} 331 (1991).

\bref J.~M. LeBlanc, J.~R. Wilson, \emph{\apj} \textbf{161} 541 (1970).

\bref Liebend\"orfer, M., Mezzacappa, A., Thielemann, F.-K., Messer, O.\,E.\,B.,
Hix, W.\,R., Bruenn, S.\,W. 2001, Phys.\ Rev.\ D 63, 103004

\bref M.~Liebend{\"o}rfer, O.~E.~B. Messer, A.~Mezzacappa, R.~W. Hix, F.-K.
  Thielemann, K.~Langanke, in \emph{Proceedings of the 11th Workshop on
  ``Nuclear Astrophysics''}, edited by W.~Hillebrandt, E.~M{\"u}ller (Ringberg
  Castle, Tegernsee, Germany, 2002), pp. 126--131.

\bref  M. Liebend\"orfer, O.~E.~B. Messer, A. Mezzacappa, S.~W. Bruenn, C.~Y. Cardall, F.-K. Thielemann, \emph{\apjs} \textbf{150} 263--316 (2004).

\bref  M. Liebend\"orfer, M. Rampp, H.-Th. Janka, A. Mezzacappa, \emph{\apj}
\textbf{620} 840--860 (2005).
 
\bref M. Liebend\"orfer, U. Pen and C. Thompson, \emph{Nuclear Physics A} \textbf{758}
  59--62 (2005).

\bref 	M. Liebend\"orfer, \emph{\apj} \textbf{633} 1042--1051 (2005).

\bref E. Livne, A. Burrows, R. Walder,
I. Lichtenstadt and T.~A. Thompson, \emph{\apj} \textbf{609} 277 (2004).

\bref A.~{Maeder}, \emph{\aap} \textbf{264}, 105--120 (1992).

\bref A.~{Maeder}, E.~K. {Grebel}, and J.-C. {Mermilliod}, \emph{\aap} \textbf{346},
  459--464 (1999).

\bref A. Maeder, G. Meynet, \emph{\aap} \textbf{440} 1041--1049 (2005).
 
\bref A. Marek, H.-Th. Janka, R. Buras, M. Liebend\"orfer, M. Rampp,
  \emph{\aap} \textbf{443} (2005).

\bref A. Marek, H. Dimmelmeier, H.-Th. Janka, E. Mueller, R. Buras, \emph{astro-ph/0502161},
  in press (2005).

\bref G. Mart\'inez-Pinedo, M. Liebend\"orfer, and D. Frekers, \emph{astro-ph/0412091}
    (2004).

\bref O.~E.~B. Messer, Ph.D. thesis, University of Tennessee (2000).

\bref G.~{Meynet}, S.~{Ekstr\"om}, and A.~{Maeder}, \emph{\aap accepted}  (2005).

\bref Nakamura, T., Umeda, H., Iwamoto, K., Nomoto, K., Hashimoto, M., Hix, W.\,R.,
Thielemann, F.-K. 2001, ApJ 555, 880

\bref J.~E. {Norris}, S.~G. {Ryan}, and T.~C. {Beers}, \emph{\apj} \textbf{561},
  1034--1059 (2001).

\bref U. Pen, P. Arras and S. Wong, \emph{\apjs} \textbf{149} 447 (2003).

\bref B.~{Plez}, and J.~G. {Cohen}, \emph{\aap} \textbf{434}, 1117--1124 (2005).

\bref N.~{Prantzos}, \emph{astro-ph/0411392, NIC8}  (2004).

\bref Pruet, J., Woosley, S.\,E., Buras, R., Janka, H.-T., Hoffman, R.\,D. 2005, ApJ
623, 325

\bref M.~Rampp, H.-T. Janka, \emph{Astron. \& Astrophys.} \textbf{396} 361 (2002).

\bref E. Ramirez-Ruiz and A. Socrates, \emph{astro-ph/0504257} (2005).

\bref Rauscher, T., Heger, A., Hoffman, R.\,D., Woosley, S.\,E. 2002, ApJ 576, 323

\bref S.~G. {Ryan}, W.~{Aoki}, J.~E. {Norris}, and T.~C. {Beers},
  \emph{astro-ph/0508475, ApJ, in press}  (2005).

\bref R.~F. Sawyer, \emph{Phys.Lett. B} \textbf{630} 1--6 (2005).

\bref  L. Scheck, T. Plewa, H.-T. Janka, K. Kifonidis, E. Mueller, \emph{Phys.Rev.Lett.}
  \textbf{92} 011103 (2004).
 
\bref S.~L. Shapiro, S.~A. Teukolsky, \emph{Black Holes White Dwarfs and Neutron
  Stars: The Physics of Compact Objects} (Wiley-Interscience, New York, 1983).

\bref M.~{Spite}, R.~{Cayrel}, B.~{Plez}, et al, \emph{\aap} \textbf{430},
  655--668 (2005).

\bref K. Sumiyoshi, S. Yamada, H. Suzuki, H. Shen, S. Chiba, H. Toki,
   \emph{\apj} \textbf{629} 922--932 (2005).

\bref  F.~D. Swesty and E~S. Myra, \emph{J.Phys.Conf.Ser.} \textbf{16} 380--389 (2005).

\bref Thielemann, F.-K., Nomoto, K., Hashimoto, M. 1996, ApJ 460, 408

\bref Thompson, C. 2000, ApJ 534, 915

\bref T.~A. Thompson, A.~Burrows, P.~A. Pinto, \emph{\apj}  \textbf{592} 434 (2003).

\bref Thompson, T.\,A., Quataert, E., Burrows, A., \emph{\apj} \textbf{620} 861--877 (2005).

\bref H.~{Umeda}, and K.~{Nomoto}, \emph{\apj} \textbf{565}, 385--404 (2002).

\bref H.~{Umeda}, and K.~{Nomoto}, \emph{\nat} \textbf{422}, 871--873 (2003).

\bref K.~A. {van Riper}, W.~D. {Arnett}, \emph{\apjs} \textbf{225}, L129 (1978).

\bref J.~S. {Vink}, and A.~{de Koter}, \emph{astro-ph/0507352}  (2005).

\bref  R. Walder, A. Burrows, C.D. Ott, E. Livne, I. Lichtenstadt, M. Jarrah,
  \emph{\apj} \textbf{626} 317--332 (2005).

\bref G.~Watanabe, K.~Sato, K.~Yasuoka, T.~Ebisuzaki, \emph{Phys. Rev. C} \textbf{69}
  055805 (2004).

\bref S.~E. Woosley, T.~A. Weaver, \emph{\apjs} \textbf{101} 181 (1995).

\bref A.~{Yahil}, \emph{\apj} \textbf{265} 1047 (1983).

}

\vfill

\end{document}